\documentclass[twocolumn,showpacs,pra,amsmath]{revtex4}
\usepackage{epsfig,graphicx,bm}

\def\be{\begin{equation}}
\def\ee{\end{equation}}
\def\e#1{\label{#1}\end{equation}}
\def\bea{\begin{eqnarray}}
\def\eea{\end{eqnarray}}
\def\ea#1{\label{#1}\end{eqnarray}}
\def\bes#1{\begin{subequations}\label{#1}}
\def\ese{\end{subequations}}

\begin{document}
\title{Error matrices in quantum process tomography}
\author{Alexander N.\ Korotkov}
\affiliation{Department of Electrical Engineering, University of
California, Riverside, California 92521}
\date{\today}

\begin{abstract}
  We discuss characterization of experimental quantum gates by the
error matrix, which is similar to the standard process matrix $\chi$
in the Pauli basis, except the desired unitary operation is factored
out, by formally placing it either before or after the error
process. The error matrix has only one large element, which is equal
to the process fidelity, while other elements are small and indicate
imperfections. The imaginary parts of the elements along the left
column and/or top row directly indicate the unitary imperfection and
can be used to find the needed correction. We discuss a relatively
simple way to calculate the error matrix for a composition of
quantum gates. Similarly, it is rather straightforward to find the
first-order contribution to the error matrix due to the
Lindblad-form decoherence. We also discuss a way to identify and
subtract the tomography procedure errors due to imperfect state
preparation and measurement. In appendices we consider several
simple examples of the process tomography and also discuss an
intuitive physical interpretation of the Lindblad-form decoherence.

    \end{abstract}




  \maketitle

\section{Introduction}

    Quantum process tomography (QPT) \cite{N-C,Poyatos-97,Chuang-97,
Leung-00,DAriano-01,Altepeter-03,Mohseni-06,Kofman-09,NJP-focus} is
a way to completely characterize a quantum process. It is the main
tool for experimental characterization of quantum gates, being
developed for potential use in a quantum computer. QPT has been
realized in numerous experiments, practically in all types of qubit
systems, including, e.g., NMR \cite{Childs-01,Weinstein-04}, linear
optics \cite{Altepeter-03,Mitchell-03,OBrien-04}, ion traps
\cite{Riebe-06,Hanneke-10}, and superconducting qubits
\cite{Neeley-08,Chow-09,Bialczak-10,Yamamoto-10,Chow-11,Dewes-12}.
In this paper we mainly focus on QPT with superconducting qubits,
even though our discussion is applicable to other systems as well.

     Unfortunately, QPT requires resources, which scale
exponentially with the number of qubits. For $N$ (superconducting)
qubits the number of initial states is usually $4^N$ (or sometimes
$6^N$), and the number of ``measurement directions'' (state
tomography ``rotations'') for each initial state is typically $3^N$
(or $6^N$). Each such setup typically requires a few hundred or a
few thousand experimental runs. From this scaling it is easy to
estimate that the QPT of a 1-qubit or a 2-qubit quantum gate
requires a manageable number of experimental runs, while a 3-qubit
QPT is rather difficult to realize, and the full QPT with 4 and more
qubits seems to be impractical.

    The problem of exponential scaling of QPT resources with
the number of qubits can be mitigated if we do not need full
information about a quantum gate operation, but instead need only
some information. Thus a partial characterization of a multi-qubit
operation is an important area of theoretical research
\cite{Emerson-05,Dankert-09,Emerson-07,Levi-07,
Bendersky-08,daSilva-11,Flammia-11}. This includes randomized
benchmarking \cite{Emerson-05,Knill-08,Magesan-11}, which typically
provides only one number: the gate fidelity. Randomized benchmarking
becomes increasingly preferable for superconducting qubits
\cite{Magesan-12,Corcoles-13,Martinis-RB}. Another promising way to
solve or a least alleviate the problem of exponential scaling is to
use a compressed-sensing implementation of  QPT
\cite{Flammia-12,Shabani-11,Rodionov-13}.

    One more problem with QPT is its sensitivity to state
preparation and measurement errors (SPAM in the terminology of Ref.\
\cite{Magesan-12}). Randomized benchmarking does not suffer from
SPAM-errors, so this is another reason why this technique is
increasingly popular. However, the obvious drawback of randomized
benchmarking is that it gives only the total (average) error and
does not give any information about particular kinds of error.
Hence, it does not tell us about the origin of a quantum gate
imperfection.

    In this paper we consider standard QPT, which gives full
information about the quantum process. Because of the scaling
problem, we are essentially talking about quantum gates with less
than 4 qubits, for which full QPT is a very useful tool. The
standard way of representing QPT results is via the process matrix
$\chi$ \cite{N-C} in the Pauli basis (see the next section).
Unfortunately, this matrix is a rather inconvenient object to work
with. Even though in principle it contains full information about
the process, it does not show useful information in a
straightforward way. Thus the important problem of converting
experimental QPT data into a useful characterization of particular
decoherence processes \cite{Boulant-03,Emerson-07,Bendersky-08,
Wolf-08,Mohseni-08,Kofman-09} is  not quite simple.

    Besides the standard matrix $\chi$, there are other ways to
represent QPT results. For example, they can be represented via the
so-called Pauli transfer matrix ${\cal R}$ \cite{Chow-12}. The
advantage of using ${\cal R}$ is that it contains only real
elements, from ${\cal R}$  it is simple to see whether the quantum
operation is trace preserving, also simple to see whether the
process is unital, and for any Clifford operation there is exactly
one non-zero element in each row and column of ${\cal R}$ with unit
magnitude.

    In this paper we discuss one more way of representing
the experimental QPT results \cite{Korotkov-unpub}. It is
essentially the standard $\chi$-matrix representation in the Pauli
basis, with the only difference being that we factor out the desired
unitary operation $U$, so that the error matrix $\chi^{\rm err}$
describes only the imperfections of the experimental quantum gate.
There are two natural ways to define such an error matrix
($\chi^{\rm err}$ and $\tilde\chi^{\rm err}$): we can assume that
the $U$-operation is either before of after the error process (see
Fig.\ 1 below).
  Even though in theoretical analyses it is rather usual
to separate the error channel and unitary operation, we are not
aware of any detailed discussion of the QPT representation by the
error matrices. As discussed in this paper, the error matrix
$\chi^{\rm err}$ (as well as $\tilde\chi^{\rm err}$) has a number of
convenient properties. In particular, its main element is equal to
the process fidelity $F_\chi$, while other non-zero elements
correspond to imperfections. Unitary imperfections are directly
given by the imaginary parts of the elements along the left column
and/or top row. Decoherence produces other elements, which have a
relatively simple relation to Kraus operators characterizing
decoherence (which are the operators in the Lindblad equation).
Since the elements of the error matrix are small, most calculations
can be approximated to first order, thus making them relatively
simple. This includes a relatively simple rule for the composition
of quantum operations and accumulation of the error-matrix elements
due to the Lindblad-form decoherence. The error-matrix
representation has already been used in the experimental QPT
\cite{Dewes-12}.

    Our paper is organized in the following way. In Sec.\
\ref{sec-standard} we briefly review some properties of the standard
process matrix $\chi$. In Sec.\ \ref{sec-chi-err} we introduce the
error matrices $\chi^{\rm err}$ and $\tilde\chi^{\rm err}$, and then
in Sec.\ \ref{sec-properties} some of their properties are
discussed, including physical interpretation. In Sec.\
\ref{sec-comp} we consider composition of the error processes. Then
in Sec.\ \ref{sec-corr} we discuss the use of the error matrix to
find the necessary unitary correction to an experimental quantum
gate. In Sec.\ \ref{sec-Lindblad-error} we consider the contribution
to the error matrix from decoherence described by the Lindblad-form
master equation. A possible identification of SPAM errors and their
subtraction from the error matrix are discussed in Sec.\
\ref{sec-SPAM}. Finally, Sec.\ \ref{sec-concl} is the conclusion.
Two appendices discuss topics that are somewhat different from the
main text. In Appendix A we consider several simple examples of the
$\chi$-matrix calculation, and in Appendix B we discuss an intuitive
interpretation of decoherence described by the Lindblad-form master
equation, unraveling the quantum dynamics into the ``jump'' and ``no
jump'' scenarios.

    \section{Standard process matrix $\chi$}
    \label{sec-standard}

A linear quantum operation $\rho_{\rm in}\rightarrow \rho_{\rm fin}$
(transforming initial density matrix $\rho_{\rm in}$ into the final
state $\rho_{\rm fin}$) is usually described via the process matrix
$\chi$ (which is Hermitian, with non-negative eigenvalues), defined
as \cite{N-C}
    \be
    \rho_{\rm fin}= \sum\nolimits_{m,n} \chi_{mn} E_m
    \rho_{\rm in}E_n^\dagger ,
    \label{chi-def}\ee
where the matrices $E_n$ form a basis in the space of complex
$d\!\times\! d$ matrices, which are the linear operators in the
$d$-dimensional Hilbert space of the problem. For example, for $N$
qubits $d=2^N$; therefore there are $(2^N)^2=4^N$ matrices $E_n$,
and the matrix $\chi$ has dimensions $4^N\! \times 4^N$. (Note that
$E_n$ are operators in the space of wavefunctions, and these
operators have the same dimension as density matrices.)

    Even though in principle any basis $\{ E_n\}$ (not necessarily
orthogonal \cite{Frobenius})  can be used in Eq.\ (\ref{chi-def}),
the most usual choice for a system of qubits is the use of Pauli
matrices. In this case for one qubit the basis $\{ E_n\}$ consists
of 4 matrices:
    \be
I\equiv\openone, \,\,\, X\equiv\sigma_x, \,\,\, Y\equiv\sigma_y,
\,\,\, Z\equiv\sigma_z,
    \ee
 and for several qubits the Kronecker product of
these matrices is used; for example for two qubits
$\{E_n\}=\{II,IX,IY, IZ,XI, ..., ZZ\}$. Note that sometimes a
different definition for $Y$ is used: $Y\equiv -i\sigma_{y}$. Also
note that the basis of Pauli matrices is orthogonal (under the
Hilbert-Schmidt inner product) but not normalized \cite{Frobenius},
so that
    \be
    \langle E_m|E_n\rangle \equiv {\rm Tr} (E_m^\dagger E_n)
    =\delta_{mn}d, \,\,\,\, d=2^N;
    \label{EmEn}\ee
    this is why normalization by $d$ in often needed in the QPT.
In this paper we will always assume the Pauli basis $\{ E_n\}$ and
use $Y\equiv \sigma_{y}$. Therefore $E_n^\dagger=E_n$, but for
clarity we will still write $E_n^\dagger$ in formulas when
appropriate.

    It is simple to find the matrix
$\chi$ for a multi-qubit unitary operation $U$. We first need to
find its represention in the Pauli basis, $U= \sum\nolimits_n u_n
E_n$,  and then compare $U\rho_{\rm in}U^\dagger$ with Eq.\
(\ref{chi-def}), which gives
 \be
 \chi_{mn}= u_m u_n^*, \,\,\, u_n =\frac{1}{d}\, {\rm
 Tr}(UE_n^\dagger),
 \label{chi-U}\ee
where the star notation means complex conjugation and the expression
for $u_n$ follows from the orthogonality property (\ref{EmEn}).
 Calculation of the matrix $\chi$ for a quantum process with
decoherence is usually significantly more cumbersome. Some examples
are considered in Appendix A (see also, e.g.,\ \cite{Kofman-09}).

    A process is called ``trace-preserving'' if the transformation
(\ref{chi-def}) preserves the trace of the density matrix, i.e.\ if
${\rm Tr}(\rho_{\rm fin})=1$ when ${\rm Tr}(\rho_{\rm in})=1$. In
this case the matrix $\chi$ should satisfy the condition
$\sum_{m,n}\chi_{mn} E_{n}^\dagger E_m=\openone$ (which gives $4^N$
real equations), and therefore $\chi$ is characterized by $16^N-4^N$
real parameters instead of $16^N$ parameters in the general
(non-trace-preserving) case. In this paper we always assume a
trace-preserving operation.

    The fidelity of a trace-preserving quantum operation compared with
a desired unitary operation is usually defined as
    \be
    F_\chi={\rm Tr} (\chi^{\rm des}\chi),
    \label{fidelity}\ee
where $\chi^{\rm des}$ is for the desired unitary operation and
$\chi$ is for the actual operation (it is easy to show that $0\leq
F_\chi\leq 1$). This definition has direct relation
\cite{Nielsen-2002,Horodecki-1999}
    \be
1-F_{\chi}=(1-F_{\rm av}) \, \frac{d+1}{d}
    \ee
 to the average state fidelity $F_{\rm av}=\overline{{\rm Tr}
 (\rho_{\rm fin}\rho_{\rm fin}^{\rm des})}$, which assumes uniform averaging
over all pure initial states. Sometimes \cite{Magesan-11} $F_{\rm
av}$ is called ``gate fidelity'' while $F_\chi$ is called ``process
fidelity''. Characterization by $F_{\rm av}$ is usually used in
randomized benchmarking
\cite{Emerson-05,Knill-08,Magesan-11,Magesan-12,Corcoles-13}; it is
easy to see that $1/(d+1)\leq F_{\rm av} \leq 1$.

    Note that the fidelity definition (\ref{fidelity}) requires unitary
desired operation and trace-preserving actual operation. For a
general (non-unitary) desired operation (still assuming
trace-preserving operations) it should be replaced with the
``Uhlmann fidelity''
    \be
F_\chi= \left({\rm Tr}\sqrt{\sqrt{\chi}\,\chi^{\rm
des}\sqrt{\chi}}\,\right)^2=\left({\rm Tr}\sqrt{\sqrt{\chi^{\rm
des}}\,\chi\sqrt{\chi^{\rm des}}}\,\right)^2 ,
    \ee
which is essentially the same definition as for the fidelity between
two density matrices \cite{N-C}.

    \section{Error matrices $\chi^{\rm err}$ and $\tilde\chi^{\rm
    err}$ } \label{sec-chi-err}

    The process matrix $\chi$ for a non-trivial unitary operation typically
has many non-zero elements (e.g., 16 elements for the two-qubit
controlled-$Z$, controlled-NOT, and $\sqrt{i{\rm SWAP}}$
operations), and it is nice-looking on the standard bar
(``cityscape'') chart used for visualization. However, a large
number of non-zero elements (we will call them ``peaks'') creates a
problem in visual comparison between the desired and experimental
$\chi$-matrices, especially for high-fidelity experiments.

    A natural way to make it easier to compare between the actual
and desired operations is to show the difference between them, i.e.,
to show the error. For example, it is possible to calculate and
display the difference $\chi-\chi^{\rm des}$ in the Pauli basis.
However, such element-by-element difference does not make much sense
mathematically.

    Instead, let us represent the actual quantum process as a
{\it composition} of the desired unitary $U$ and the error process
(Fig.\ 1), and find the process matrix $\chi^{\rm err}$ for this
error operation. This essentially reduces the comparison between
$\chi$ and desired $U$ (we use a loose language here) to the
comparison between $\chi^{\rm err}$ and the memory (identity)
operation.

\begin{figure}[tb]
  \centering
\includegraphics[width=8.5cm]{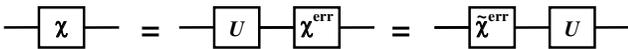}
 \vspace{-0.2cm}
  \caption{Quantum circuit diagrams, which define the error processes
characterized by error matrices $\chi^{\rm err}$ and
$\tilde\chi^{\rm err}$ via their relation to $\chi$ and the desired
unitary operation $U$. The time runs from left to right.}
  \label{Idea}
\end{figure}

    So, the general idea is to convert the desired unitary $U$ into the
memory operation, and this converts $\chi$ into the error process
matrix $\chi^{\rm err}$.
    There are two ways of defining such error matrix: we can
place the error process either before or after the desired unitary
$U$ (Fig.\ 1). Thus, we define two error matrices: $\chi^{\rm err}$
and $\tilde\chi^{\rm err}$ using the relations [see Eq.\
(\ref{chi-def})]
     \begin{eqnarray}
&&   \rho_{\rm fin}= \sum\nolimits_{m,n} \chi_{mn}^{\rm err}
   E_m U \rho_{\rm in}U^\dagger E_n^\dagger ,
    \label{chi-err-def}\
        \\
&&   \rho_{\rm fin}= \sum\nolimits_{m,n} \tilde\chi_{mn}^{\rm err}
   U E_m \rho_{\rm in}E_n^\dagger U^\dagger .
    \label{tilde-chi-err-def}\end{eqnarray}

    From Fig.\ 1 it is obvious that the process $\chi^{\rm err}$ can
be represented as the composition of the inverse ideal unitary
($U^{-1}=U^\dagger$) and the actual process $\chi$ after that.
Similarly, $\tilde\chi^{\rm err}$ is the composition of $\chi$ and
inverse unitary $U^\dagger$ after that. Hence, $\chi^{\rm err}$ and
$\tilde\chi^{\rm err}$ represent legitimate quantum processes and
therefore they satisfy all the properties of the usual matrix $\chi$
(see the previous section); in particular, $\chi^{\rm err}$ and
$\tilde\chi^{\rm err}$ are positive Hermitian matrices. We will use
the standard Pauli basis for the error matrices $\chi^{\rm err}$ and
$\tilde\chi^{\rm err}$.

    Using the composition relation, the
error matrices can be calculated from $\chi$ as \cite{Kofman-09}
    \begin{eqnarray}
  \chi^{\rm err}= V\chi V^\dagger, \,\,\,
  V_{mn}={\rm Tr} (E^\dagger_m E_n U^\dagger) /d ,
    \label{chi-err-chi}    \\
  \tilde\chi^{\rm err}= \tilde{V}\chi \tilde{V}^\dagger, \,\,\,
  \tilde{V}_{mn}={\rm Tr} (E^\dagger_m U^\dagger E_n ) /d .
    \label{chi-tilde-err-chi}\end{eqnarray}
    In an experiment the error process matrices can be calculated
either from $\chi$ using these equations or by directly applying the
definitions (\ref{chi-err-def}) and (\ref{tilde-chi-err-def}) to the
experimental data. For example, to find $\tilde\chi^{\rm err}$ the
measured final density matrices can be transformed numerically as
$\rho_{\rm fin}\rightarrow U^\dagger\rho_{\rm fin}U$ and then the
usual procedure of $\chi$ calculation can be applied. Note that the
matrix $\chi^{\rm err}$ can be thought of as the $\chi$-matrix in
the rotated basis $\{E_n U\}$ instead of $\{ E_n\}$; similarly,
$\tilde\chi^{\rm err}$ is formally the $\chi$-matrix in the basis
$\{U E_n\}$; however, we will not use this language to avoid
possible confusion.

    In the ideal (desired) case both error processes $\chi^{\rm err}$
and $\tilde\chi^{\rm err}$ are equal to the perfect memory
(identity) operation, $\chi^{\rm err}=\tilde\chi^{\rm err}=\chi^{\bf
I}$, for which
   \be
    \chi^{\bf I}_{mn}=\delta_{m0}\delta_{n0},
    \label{memory}\ee
where with index 0 we denote the left column and/or the top row,
which correspond to the unity basis element (in the usual notation
$0=I$ for one qubit, $0=II$ for two qubits, etc.); the process
matrix (\ref{memory}) is given by Eq.\ (\ref{chi-U}) with
$u_n=\delta_{n0}$.
  So, in the ideal case the error matrices have
only one non-zero element at the top left corner: $\chi^{\rm
err}_{00}=\tilde\chi^{\rm err}_{00}=1$.   Therefore, {\it any other
non-zero element in $\chi^{\rm err}$ (or $\tilde\chi^{\rm err}$)
indicates an imperfection of the quantum operation}. This is the
main advantage in working with $\chi^{\rm err}$ or $\tilde\chi^{\rm
err}$ instead of the usual matrix $\chi$. There are also some other
advantages discussed below.

    The standard process fidelity (\ref{fidelity}) for a
trace-preserving operation has a very simple form for the error
matrices. Since $\chi$, $\chi^{\rm err}$, and $\tilde\chi^{\rm err}$
are essentially the same operator in different bases, we have ${\rm
Tr} (\chi^{\rm des} \chi )={\rm Tr} (\chi^{\bf I} \chi^{\rm
err})={\rm Tr} (\chi^{\bf I} \tilde\chi^{\rm err})$. Therefore
    \be
    F_\chi = \chi_{00}^{\rm err}= \tilde\chi_{00}^{\rm err},
    \label{F-chi-error}\ee
i.e.\ the process fidelity is just the height of the main (top left)
element of the error matrix $\chi^{\rm err}$ or $\tilde\chi^{\rm
err}$.

    A systematic unitary error can be easily detected (to first order)
in the error matrix $\chi^{\rm err}$ or $\tilde\chi^{\rm err}$
because it appears at a special location: it produces non-zero
imaginary elements along the top row and left column of the matrix,
i.e.\ the elements ${\rm Im}(\chi^{\rm err}_{0n})=-{\rm
Im(}\chi^{\rm err}_{n0})$ with $n\neq 0$ (similarly for
$\tilde\chi^{\rm err}$). To see this fact, let us assume that
instead of the desired unitary operation $U$, the quantum gate
actually realizes a slightly imperfect unitary $U^{\rm actual}$.
Then $\chi^{\rm err}$ corresponds to the unitary $U^{\rm err}=U^{\rm
actual} U^{\dagger}$, which can be expanded in the Pauli basis as
$U^{\rm err}=\sum_n u^{\rm err}_n E_n$. Since $U^{\rm err}\approx
\openone$, we have $u^{\rm err}_0\approx 1$ and $|u^{\rm err}_{n\neq
0}|\ll 1$. Note that $u^{\rm err}_0$ can always be chosen real
because $U^{\rm err}$ is defined up to an overall phase. Now let us
show that to first order all $u^{\rm err}_{n\neq 0}$ are purely
imaginary. This follows from the first-order expansion of the
equation $U^{\rm err}U^{{\rm err}\,\dagger}=\openone$, which gives
$|u_0^{\rm err}|^2\openone + \sum_{n\neq 0} (u_n^{\rm err}+u_n^{{\rm
err}\,*}) E_n=\openone $. Hence, $u_n^{\rm err}+u_n^{{\rm
err}\,*}=0$ for $n\neq 0$ (purely imaginary $u^{\rm err}_{n\neq
0}$), and the difference $u^{\rm err}_0 -1$ is only of second order
(for a real $u^{\rm err}_0$). Another way of showing that in the
first order $u^{\rm err}_{n\neq 0}$ are imaginary is by using
representation $U^{\rm err}=e^{iH_{\rm err}}$ (neglecting the
overall phase) with a Hermitian matrix $H_{\rm err}$, so that the
expansion $H_{\rm err}=\sum_n h^{\rm err}_n E_n$ contains all real
coefficients, while to first order $u^{\rm err}_{n\neq 0}=ih^{\rm
err}_{n\neq 0}$ and $u^{\rm err}_0=1$.

 Now using Eq.\ (\ref{chi-U}) we see that to first order
the only non-zero elements (except $\chi_{00}^{\rm err}$) are
    \be
    {\rm Im}(\chi^{\rm err}_{n0})=-{\rm Im}(\chi^{\rm
    err}_{0n})\approx
    -i u^{\rm err}_{n}, \,\,\, n\neq 0.
    \label{imag-unitary}\ee
Note that the diagonal elements $\chi^{\rm err}_{nn}$ (as well as
the change of $\chi_{00}^{\rm err}$) in this case are of  second
order. In particular, in this approximation $F_\chi \approx 1$. As
discussed later, if decoherence causes a significant decrease of the
fidelity $F_\chi$, then a better approximation of the systematic
unitary error effect is Eq.\ (\ref{imag-unitary}) multiplied by the
fidelity,
    \be
{\rm Im}(\chi^{\rm err}_{n0})\approx  -i F_\chi u^{\rm err}_{n}.
    \label{imag-unitary-2}\ee
 Using this
equation it is possible to find $u_n^{\rm err}$ from an experimental
matrix $\chi^{\rm err}$ and therefore estimate the systematic
unitary error $U^{\rm err}$ in the experiment.

 The same property (\ref{imag-unitary}) [and its version
(\ref{imag-unitary-2}) corrected for $F_\chi$] can be shown for
$\tilde\chi^{\rm err}$ using the similar derivation for the unitary
imperfection $\tilde{U}^{\rm err}=U^{\dagger}U^{\rm actual}\approx
\openone$. The elements  ${\rm Im}(\tilde\chi^{\rm err}_{n0})\approx
-i F_\chi \tilde{u}^{\rm err}_{n}$ are different compared with the
matrix elements of $\chi^{\rm err}$; they are related via the
equation $\tilde{U}^{\rm err}=U^\dagger U^{\rm err}U$ or equivalent
equation $\tilde{U}^{\rm err}-\openone =U^\dagger (U^{\rm
err}-\openone )U$.

    Decoherence produces additional small peaks in the error matrix
$\chi^{\rm err}$ (and/or $\tilde\chi^{\rm err}$). As discussed
later, to first order these peaks are linear in the decoherence
strength and simply additive for different decoherence mechanisms.
Therefore, to first order we have a weighted sum of different
patterns in $\chi^{\rm err}$ for different mechanisms. If these
patterns for the most common decoherence mechanisms are relatively
simple, then there is a rather straightforward way of extracting
information on decoherence from experimental $\chi^{\rm err}$
matrix. In Sec.\ \ref{sec-Lindblad-error} we will discuss the
general way to calculate the first-order pattern for a particular
decoherence mechanism; for a practical quantum gate this pattern may
contain many elements. A special role is played by the real elements
along the top row and left column of $\chi^{\rm err}$ (or
$\tilde\chi^{\rm err}$): they correspond to the gradual non-unitary
(``Bayesian'') evolution in the absence of the ``jumps'' due to
decoherence (in contrast to the imaginary elements, which correspond
to the unitary imperfection) -- see discussion later.

    Note that the diagonal matrix elements of $\chi^{\rm err}$
are the error probabilities in the so-called Pauli twirling
approximation  \cite{Emerson-05,Lopez-10,Ghosh-12,Geller-13}.
Therefore these elements can be used in simulation codes, which use
the Pauli twirling approximation for the analysis of quantum
algorithms in multi-qubit systems.

    Concluding this section, we emphasize that the error matrix
$\chi^{\rm err}$ (and/or $\tilde\chi^{\rm err}$) is just a minor
modification of the standard $\chi$ matrix; they are related by a
linear transformation and therefore equivalent to each other.
However, in the error matrix only one peak (at the top left corner)
corresponds to the desired operation, while other peaks indicate
imperfections. This makes the error matrix more convenient to work
with, when we analyze deviations of an experimental quantum
operation from a desired unitary  and try to extract information
about the main decoherence mechanisms.

    \section{Some properties of the error matrices and interpretation}
    \label{sec-properties}

    In this paper we always assume high-fidelity operations,
    \be
    1-F_\chi \ll 1 ,
    \label{high-fid}\ee
so that the first-order approximation of imperfections is quite
accurate.
      Since the error matrix $\chi^{\rm err}$ is positive, its
off-diagonal elements have the upper bound
    \be
|\chi^{\rm err}_{mn}| \leq \sqrt{\chi^{\rm err}_{mm}\chi^{\rm
err}_{nn}}.
    \ee
(The same is true for $\tilde\chi^{\rm err}$, but for brevity we
discuss here only $\chi^{\rm err}$.) Therefore for a high-fidelity
operation (\ref{high-fid}) only the elements in the left column and
top row can be relatively large, $|\chi^{\rm err}_{0n}|= |\chi^{\rm
err}_{n0}|\leq \sqrt{\chi^{\rm err}_{nn}}\leq \sqrt{1-F_\chi}$,
while other off-diagonal element have a smaller upper bound,
$|\chi^{\rm err}_{mn}|\leq (1-F_\chi)/2$, because all diagonal
elements except $\chi^{\rm err}_{00}$ are small. Actually, it is
possible to show (see below) that ${\rm Re}(\chi^{\rm err}_{0n})$
are also small, $|{\rm Re}(\chi^{\rm err}_{0n})|\leq (1-F_\chi )/2$,
so only ${\rm Im}(\chi^{\rm err}_{0n})$ can be relatively large.

    The elements ${\rm Im}(\chi^{\rm err}_{0n})$ and
${\rm Im}(\chi^{\rm err}_{n0})$ play a special role because a
unitary imperfection produce them in the first order, while other
elements are of second order [see Eq.\ (\ref{chi-U}) and discussion
in the previous section]. It is easy to see that ${\rm Im}(\chi^{\rm
err}_{0n})$ and ${\rm Im}(\chi^{\rm err}_{n0})$ can approach
$\pm\sqrt{\chi_{nn}^{\rm err}}$ if the error is dominated by the
unitary imperfection.

    The physical intuition in analyzing the error of a quantum gate
is that the (small) infidelity $1-F_\chi$ comes from small unitary
imperfections and from rare but ``strong'' decoherence processes,
which cause ``jumps'' that significantly change the state.
   As discussed later, this picture should also be complemented by
small non-unitary state change in the case when no jump occurred
(this change is essentially the quantum Bayesian update
\cite{Molmer-92,Kor-Bayes,Katz-06} due to the information that there
was no jump).

      It is not easy to formalize this intuition mathematically;
however, there is a closely related procedure. Let us diagonalize
the error matrix $\chi^{\rm err}$, so that
    \be
    \chi^{\rm err}= TDT^{-1},
    \label{diag-chi-err}\ee
where $T$ is a unitary $d^2\!\times d^2$ matrix and $D={\rm
diag}(\lambda_0, \lambda_1,\dots)$ is the diagonal matrix containing
$d^2$ real eigenvalues of $\chi^{\rm err}$. These eigenvalues are
non-negative, and we can always choose $T$ so that in $D$ they are
ordered as $\lambda_0\geq\lambda_1\geq \dots  \geq \lambda_{d^2-1}$.
The sum of the eigenvalues is equal 1 (since we consider
trace-preserving operations) and the largest eigenvalue $\lambda_0$
is close to 1 because $F_\chi\leq \lambda_0\leq 1$. [Here
$\lambda_0\geq \chi^{\rm err}_{00}$ because any diagonal element of
a Hermitian matrix should be in between the largest and smallest
eigenvalues, as follows from expanding the corresponding basis
vector in the eigenbasis.] All other eigenvalues are small because
$\sum_{k>0}\lambda_k \leq 1-F_\chi$.

    The diagonalization (\ref{diag-chi-err}) directly gives the
evolution representation via the Kraus operators \cite{Kraus-book},
    \begin{eqnarray}
&& \rho_{fin} = \sum_{k=0}^{d^2-1} \lambda_k A_k
(U\rho_{in}U^\dagger) A_k^\dagger,  \,\,\, \sum_k \lambda_k
A_k^\dagger A_k =\openone , \qquad
    \label{Kraus-representation} \\
&& A_k=\sum_n a_n^{(k)}E_n, \,\,\, a_n^{(k)}=T_{nk},
    \label{Kraus-representation-2}\end{eqnarray}
in which the operators $A_k$ form an orthogonal basis, $\langle
A_m|A_n\rangle =\delta_{mn}d$ [see Eq.\ (\ref{EmEn})] because $T$ is
unitary. Recall that $U$ is the desired unitary gate -- see Eq.\
(\ref{chi-err-def}). The main term in the Kraus-operator
representation (\ref{Kraus-representation}) is the term with
$\lambda_0\approx 1$; let us show that $A_0\approx \openone$ (up to
an overall phase, which can always be eliminated); this means
$a_0^{(0)}\approx 1$ and $|a_n^{(0)}|\ll 1$ for $n\neq 0$ [note that
$\sum_n |a_n^{(0)}|^2=1$ because $T$ is unitary; therefore it is
sufficient to show that $a_0^{(0)}\approx 1$]. This can be done in
the following way. Since
    \be
\chi^{\rm err}_{mn}=\sum\nolimits_k \lambda_k
a_m^{(k)}(a_n^{(k)})^*,
    \label{chi-via-a}\ee
the fidelity is
    \be
    F_\chi=\sum\nolimits_k\lambda_k |a_0^{(k)}|^2.
    \label{F-sum-k}\ee
     The contribution of the terms with $k\neq 0$ to
the fidelity is at least smaller than $1-F_\chi$ because this is the
bound for $\sum_{k>0}\lambda_k$ and $|a_0^{(k)}|=|T_{0k}|< 1$ for a
unitary $T$ (this contribution is actually even much smaller, as
discussed below). Therefore $\lambda_0 |a_0^{(0)}|^2
>2F_\chi -1$, and so $|a_0^{(0)}|^2>1-2(1-F_\chi )$. Choosing real
$a_0^{(0)}$, we obtain $a_0^{(0)}\approx 1$, and therefore
$A_0\approx E_0= \openone$.

    Since $A_0$ is close to $E_0$ and other $A_k$ are orthogonal to
$A_0$, the components $a_0^{(k)}$ are small for $k\neq 0$. Using the
relation $(a_0^{(0)})^* a_0^{(k)} + \sum_{n>0} (a_n^{(0)})^*
a_n^{(k)}=0$ (since $T$ is unitary), we find the bound $|a_0^{(k)}|
< \sqrt{\sum_{n>0} |a_n^{(0)}|^2 } \sqrt{\sum_{n>0} |a_n^{(k)}|^2
}/|a_0^{(0)}|$. Now using $|a_0^{(0)}|\approx 1$, $\sum_{n>0}
|a_n^{(k)}|^2 \leq 1$, and $\sum_{n>0} |a_n^{(0)}|^2 <2(1-F_\chi)$
(see the previous paragraph), we obtain $|a_0^{(k)}|^2 <
2(1-F_\chi)$, neglecting the terms of the order $(1-F_\chi)^3$. This
means that the contribution to the fidelity  (\ref{F-sum-k}) from
the Kraus operators with $k>0$ is limited at least by
$2(1-F_\chi)^2$ [neglecting the order $(1-F_\chi)^4$], and therefore
a good approximation for the fidelity is
    \be
    F_\chi\approx \lambda_0 |a_0^{(0)}|^2, \,\,\,
    1-F_\chi \approx (1-|a_0^{(0)}|^2 ) + (1-\lambda_0),
    \label{F-separation}\ee
which corresponds to the intuitive separation of the infidelity
$1-F_\chi$ into the ``coherent'' error $1-|a_0^{(0)}|^2$ and the
error $1-\lambda_0$ due to rare but strong decoherence ``jumps''.

    Thus the Kraus-operator representation (\ref{Kraus-representation})
can be crudely interpreted in the following way.
   After the desired unitary $U$, we
apply the Kraus operator $A_0\approx \openone$ with the probability
$\lambda_0\approx 1$, while with small remaining probabilities
$\lambda_k$ we apply very different (orthogonal to $A_0$) Kraus
operators $A_k$. Imperfection of $A_0$ (compared with $\openone$)
leads to the ``coherent'' error $1-|a_0^{(0)}|^2$ with
$a_0^{(0)}={\rm Tr} (A_0)/d$. Other Kraus operators are practically
orthogonal to $E_0=\openone$, so they correspond to ``strong
decoherence'' and practically always lead to an error, which happens
with the total probability $\sum_k \lambda_k=1-\lambda_0$, thus
explaining Eq.\ (\ref{F-separation}). While this interpretation
seems to be quite intuitive, there are two caveats.
   First, it is incorrect to say that $A_k$ is applied with
the probability $\lambda_k$. Instead, we should say
\cite{N-C,Kraus-book} that the evolution scenario
    \be
    |\psi_{\rm in}\rangle \rightarrow \frac{A_k U |\psi_{\rm in}\rangle}
{\rm Norm},     \,\,\, \rho_{\rm in} \rightarrow \frac{A_kU\rho_{\rm
in}U^\dagger A_k^\dagger}{\rm Norm}
    \ee
occurs with the probability $P_k=\lambda_k {\rm Tr}(A_k^\dagger A_k
U\rho_{\rm in}U^\dagger)$, which depends on the initial state and is
equal to $\lambda_k$ only on average, $\overline{P_k}=\lambda_k$,
after averaging over pure initial states. [This can be proven by
using $\overline{\rho_{\rm in}}=\openone /d$ and ${\rm
Tr}(A_k^\dagger A_k)=d$.] Note that the operators $A_k$ can violate
the inequality $A_k^\dagger A_k \leq \openone$, even though
$\lambda_k A_k^\dagger A_k \leq \openone$ is always satisfied. The
second caveat is that $A_0$ is not necessarily unitary, as would be
naively expected for the separation into a coherent operation and
decoherence. The non-unitary part of $A_0$ can be interpreted as due
to the absence of jumps $A_{k>0}$, similar to the null-result
evolution \cite{Molmer-92,Katz-06} (see below and also discussions
in Sec.\ \ref{sec-Lindblad-error} and Appendix B). We may include
both contributions (imperfect unitary part and non-unitary part of
$A_0$) into what we call the ``coherent error'', so it is
characterized by the difference between $A_0$ and $\openone$
(probably it is better to call it the ``gradual error''). With
understanding of these two caveats, the discussed above
interpretation of the Kraus representation
(\ref{Kraus-representation}) can be useful for gaining some
intuition.

    Note that the contribution to $\chi^{\rm err}$ [see Eq.\
(\ref{chi-via-a})] from the imperfection of the main Kraus operator
$A_0$ mainly causes the elements $\chi_{n0}^{\rm
err}=(\chi_{0n}^{\rm err})^*\approx \lambda_0 a_n^{(0)}$ in the top
row and left column of the error matrix (other elements are of the
second order in the imperfection). In contrast, other (``decoherence
jump'') operators $A_k$ mainly produce other elements of $\chi^{\rm
err}$; their contribution to $\chi^{\rm err}_{0n}$ is limited by
$\sqrt{2}(1-F_\chi)\sqrt{\chi_{nn}^{\rm err}}$, as follows from the
positivity of $\chi_{mn}^{\rm dec}\equiv \sum_{k>0}\lambda_k
a_m^{(k)}(a_n^{(k)})^*$, which gives $|\chi_{0n}^{\rm dec}|\leq
\sqrt{\chi_{00}^{\rm dec}\chi_{nn}^{\rm dec}}$, and the derived
above inequality $\chi_{00}^{\rm dec}\leq 2(1-F_\chi)^2$.
Significant contributions to the diagonal elements of $\chi^{\rm
err}$ may come from both $A_0$ and $A_{k>0}$.

    Therefore, we can apply the following approximate procedure
to crudely separate the error matrix,
    \be
\chi^{\rm err}=\chi^{\rm coh} + \chi^{\rm dec}, \,\,\, \chi^{\rm
coh}_{mn}= \lambda_0 a_m^{(0)}(a_n^{(0)})^*,
    \label{chi-coh-dec}\ee
into the ``coherent'' (or ``gradual'') part $\lambda_0
a_m^{(0)}(a_n^{(0)})^*$ and ``strong decoherence'' $\chi^{\rm dec}$
[see Eq.\ (\ref{chi-via-a})]. We first estimate the ``coherent
probability'' $\lambda_0$ as
    \be
    \lambda_0 \approx F_\chi / \left( 1-\sum\nolimits_{n>0}
    |\chi_{0n}^{\rm err}|^2\right) ,
    \ee
and then use this $\lambda_0$ in the estimation
$a_{n>0}^{(0)}\approx \chi^{\rm err}_{n0}/\lambda_0$. Using
$\lambda_0$ and $a_n^{(0)}$ we can construct $\chi^{\rm coh}$, while
the remaining part of $\chi^{\rm err}$ is $\chi^{\rm dec}$. The
diagonal elements $\chi^{\rm err}_{nn}$ with $n\neq 0$ (error
probabilities) are thus separated into the contributions $\lambda_0
|a_n^{(0)}|^2$ from the ``coherent imperfection'' and
$\chi_{nn}^{\rm err}-\lambda_0 |a_n^{(0)}|^2$ due to ``strong
decoherence''. A further simplification of this procedure is to
approximate the coherent part as $\chi^{\rm coh}_{mn}\approx
\chi^{\rm err}_{m0}(\chi_{0n}^{\rm err})^*$ for $m\neq 0$, $n\neq
0$, and $\chi^{\rm coh}_{m0}=(\chi^{\rm coh}_{0m})^*\approx
\chi^{\rm err}_{m0}$.

    One more useful approach for the intuitive understanding of
$\chi^{\rm err}$ elements along the left column and top row is the
following. Using the completeness relation
(\ref{Kraus-representation}) we can write the main (``coherent'')
Kraus operator in the polar decomposition representation as
    \begin{eqnarray}
&&    \sqrt{\lambda_0} \, A_0 =U^{\rm err} \sqrt{\openone -
    \sum\nolimits_{k>0} \lambda_k A_k^\dagger A_k }
        \label{A0-pd}\\
&& \hspace{1cm} \approx U^{\rm err} \left( \openone -\frac{1}{2}
\sum\nolimits_{k>0} \lambda_k A_k^\dagger A_k \right) ,
    \end{eqnarray}
where $U^{\rm err}\approx \openone$ is some unitary, which
corresponds to the unitary imperfection [since the overall phase is
arbitrary, we can choose ${\rm Im}({\rm Tr} (U^{\rm err}))=0$]. Then
let us expand the operators in the Pauli basis,
    \begin{eqnarray}
    U^{\rm err} \approx (1-\frac{1}{2}{\cal E}_U)E_0 +\sum_{n>0} u^{\rm
    err}_n E_n, \,\, {\cal E}_U\equiv \sum_{n>0} |u_n^{\rm
    err}|^2, \quad
    \label{expansion-unitary} \\
    \sum_{k>0} \lambda_k A_k^\dagger A_k = {\cal E}_D E_0 +
    \sum_{n>0} g_n E_n, \,\,\, {\cal E}_D\equiv 1-\lambda_0, \qquad
    \label{expansion-dec}\end{eqnarray}
where $u_n^{\rm err}$ and $g_n$ are the expansion components, and we
introduced the naturally defined unitary error ${\cal E}_U$ and
decoherence error ${\cal E}_D$ (average probability of ``jumps'').
Now using Eq.\ (\ref{F-separation}) and neglecting the second-order
products $u^{\rm err}_m g_n$, we find the intuitively expected
formula for fidelity,
    \be
    F_\chi \approx 1 - {\cal E}_U -{\cal E}_D.
    \ee
Similarly, taking into account only the contribution from the main
Kraus operator, for the elements $\chi^{\rm err}_{n0}=\chi^{{\rm
err}*}_{0n}$ with $n\neq 0$ we find
    \be
    \chi^{\rm err}_{n0} \approx (1-\frac{1}{2}{\cal E}_U-{\cal
    E}_D)\, u^{\rm err}_n + (1-{\cal E}_U-\frac{1}{2}{\cal
    E}_D)\,\frac{g_n}{2}.
    \label{chi-n0-approx}\ee
Here $u^{\rm err}_n$ are purely imaginary (in the first order)
because $U^{\rm err}$ is unitary, while $g_n$ are real because Eq.\
(\ref{expansion-dec}) is the expansion of a Hermitian operator.
Therefore, we see that the imaginary parts of the elements
$\chi^{\rm err}_{n0}$ are due to unitary imperfection, while their
real parts come from the absence of ``jumps'' (described by Kraus
operators with $k>0$) via Eq.\ (\ref{A0-pd}). It is easy to see that
the evolution $\sqrt{\openone -\sum\nolimits_{k>0} \lambda_k
A_k^\dagger A_k }$ is essentially the Bayesian update of the quantum
state \cite{Molmer-92,Kor-Bayes,Katz-06} due the absence of jumps.

    In experiments with superconducting qubits the quantum gate
infidelity is usually dominated by decoherence,  ${\cal E}_D \gg
{\cal E}_U$, unless the unitary part is very inaccurate. We will
often assume this situation implicitly. In this case case $F_\chi
\approx 1- {\cal E}_D$, and from Eq.\ (\ref{chi-n0-approx}) we
obtain Eq.\ (\ref{imag-unitary-2}).

    Using Eqs.\ (\ref{expansion-dec}) and (\ref{chi-n0-approx}) we
can show the bound  $|{\rm Re} (\chi_{n0}^{\rm err})| \leq
(1-\lambda_0)/2\leq (1-F_\chi)/2$. The starting point is to see that
all components of $A_k^\dagger A_k$ in the Pauli basis are not
larger than 1, i.e.\ $|{\rm Tr} (E^\dagger_n A_k^\dagger A_k)/d|\leq
1$ for any $n$. This is because $A_k^\dagger A_k=\sum_{m,l}
(a_m^{(k)})^*E_m^\dagger a_l^{(k)}E_l$ and the product of two Pauli
operators is a Pauli operator (with a phase factor); therefore a
particular (say, $n$th) component  is essentially a sum of pairwise
products of the ``vector coordinates'' $(a^{(k)}_m)^*$ and the same
coordinates $a^{(k)}_l$ in a different order (also, with phase
factors). Therefore, the sum of products is limited by the norms of
the two vectors, which is 1 for both vectors ($\sum_m
|a_n^{(k)}|^2=1$). Since the Pauli-basis components of  $A_k^\dagger
A_k$ are limited by 1, the sum of $k$th contributions to $g_n$ in
Eq.\ (\ref{expansion-dec}) is limited by $|g_n|\leq
\sum_{k>0}\lambda_k=1-\lambda_0$. This gives $|{\rm Re}(\chi^{\rm
err}_{n0})|\leq (1-\lambda_0)/2$ via Eq.\ (\ref{chi-n0-approx}).

    Since the elements ${\rm Re} (\chi^{\rm err}_{n0})$ are small,
so that their second-order contributions to the diagonal elements
are practically negligible, $[{\rm Re} (\chi^{\rm err}_{n0})]^2\leq
(1-F_\chi)^2/4$, it does not matter much whether we include the
evolution due to the ``no-jump'' Bayesian update into the ``coherent
part'' (\ref{chi-coh-dec}) or not. Note that this Bayesian evolution
does not produce an additional error since ${\rm Tr}[E_0^\dagger
(1-\frac{1}{2}\sum_{k>0}\lambda_k A_k^\dagger
A_k)]/d=1-\frac{1}{2}\sum_{k>0}\lambda_k=1-(1-\lambda_0)/2$ and
therefore from Eq.\ (\ref{A0-pd}) we see that $a^{(0)}_0 \approx
1-\frac{1}{2}\sum_{k>0}|u_k^{\rm err}|^2$, i.e.\ the ``no-jump''
scenario brings only the unitary error.

\vspace{0.3cm}

    In this section we discussed only the error matrix
$\chi^{\rm err}$; however, the same analysis can be also applied to
$\tilde\chi^{\rm err}$.

  The relation between the error matrices is \cite{Kofman-09}
  \be
  \chi^{\rm err}=W_{(U)}\tilde\chi^{\rm err}W_{(U)}^\dagger, \,\,\,
  \tilde\chi^{\rm err} = W_{(U)}^\dagger \chi^{\rm err} W_{(U)},
  \label{W-transf}\ee
where the unitary matrix $W_{(U)}$ corresponds to the effective
change of the basis $\{E_n\}$ due to $U$,
    \be
W_{(U),mn}= {\rm Tr} (E_m^\dagger U E_n U^\dagger)/d.
    \label{W-def}\ee
Note that $W^\dagger_{(U)}=W_{(U^\dagger)}$. It is convenient to
think that the $W$-transformation (\ref{W-transf}) is due to the
error matrix ``jumping over'' the unitary $U$ (see Fig.\ 1). It is
easy to see from Eq.\ (\ref{W-def})  that $W_{0n}=\delta_{0n}$;
therefore for an ideal memory $W\chi^{\bf I}W^\dagger =\chi^{\bf I}$
and Eq.\ (\ref{W-transf}) essentially transforms only the difference
from the ideal memory.

    The $W^\dagger$-transformation $\chi^{\rm err}\rightarrow
    \tilde\chi^{\rm err}$ (\ref{W-transf}) obviously corresponds to
the unitary transformation of Kraus operators $A_k$, ``jumping
over'' $U$ to the left (Fig.\ 1),
    \be
    \tilde{A}_k= U^\dagger A_k U .
    \ee
These $\tilde A_k$ form the Kraus-operator representation of
$\tilde\chi^{\rm err}$ with the same ``probabilities'' $\lambda_k$.

    \section{Composition of error processes }
    \label{sec-comp}

    Let us calculate the error matrix $\chi^{\rm err}$ for the
composition of two quantum operations: desired unitary $U_1$ with
error process $\chi_1^{\rm err}$ and after that the desired unitary
$U_2$ with error $\chi^{\rm err}_2$ (Fig.\ 2). It is obvious that
the resulting desired unitary is $U_2U_1$ (note that the matrix
multiplication is from right to left, while on the quantum circuit
diagrams the time runs from left to right). We assume sufficiently
high fidelity of both operations, $F_1\simeq 1$, $F_2\simeq 1$  [for
brevity we omit the subscript $\chi$ in the notation
(\ref{F-chi-error}) for fidelity].

\begin{figure}[tb]
  \centering
\includegraphics[width=8.5cm]{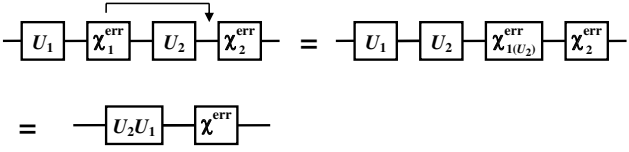}
 \vspace{-0.2cm}
  \caption{Quantum circuit diagram for the composition of two
quantum operations.  The arrow illustrates ``jumping'' the error
process $\chi_1^{\rm err}$ over the unitary $U_2$.}
  \label{fig-composition}
\end{figure}

    Let us start with the simple case when there are no unitaries,
$U_1=U_2=\openone$. Then the relation between the initial state
$\rho_{\rm in}$ and the final state $\rho_{\rm fin}$ is
    \be
    \rho_{\rm fin} = \sum_{m,n,p,q} \chi^{\rm err}_{1, mn}
\chi^{\rm err}_{2, pq} E_p E_m  \rho_{\rm in} E_n^\dagger
E_q^\dagger,
    \ee
which leads to the usual lengthy expression for the composition of
two operations:
    \be
    \chi^{\rm err}_{ab}=  \sum_{m,n,p,q} \chi^{\rm err}_{1, mn}
\chi^{\rm err}_{2, pq} \frac{1}{d^2} {\rm Tr} (E_p E_m E_a^\dagger)
{\rm Tr} (E_q E_n E_b^\dagger)^*,
    \label{comp-gen}\ee
where we used relation
    \be
    E_p E_m= \sum_a \frac{1}{d} {\rm Tr} (E_p E_m E_a^\dagger) E_a
    \ee
(actually there is only one non-zero term in this relation, but
there is no simple way to write it). Equation (\ref{comp-gen}) is
valid for any two operations (not necessarily error processes) and
is very inconvenient to use. Fortunately, it is greatly simplified
for the error processes, since $\chi_{1}^{\rm err}$ and
$\chi_{2}^{\rm err}$ contain only one large (close to one) element,
$\chi_{1,00}^{\rm err}=F_1$ and $\chi_{2,00}^{\rm err}=F_2$, while
all other elements are small. Therefore we can use the first order
approximation of Eq.\ (\ref{comp-gen}), which gives the simple
additive relation
    \begin{eqnarray}
&&  \chi_{mn}^{\rm err}  \approx  F_2   \chi_{1,mn}^{\rm err}+
    F_1 \chi_{2,mn}^{\rm err}
    \label{comp-err}\\
&&\hspace{0.75cm}     \approx  \chi_{1,mn}^{\rm err}+
    \chi_{2,mn}^{\rm err}
    \label{comp-err-2}\end{eqnarray}
for all elements except the main element $\chi_{00}^{\rm err}$. The
approximation (\ref{comp-err-2}) can also be written as
    \be
 \chi^{\rm err} \approx \chi_{1}^{\rm err}+
    \chi_{2}^{\rm err}-\chi^{\bf I}.
    \label{comp-err-matrix}\ee
 Obviously, it is much easier to deal with the composition of
error processes than with the composition of general quantum
processes. Note that Eq.\ (\ref{comp-err}) can also be naturally
understood using the Kraus-operator representation
(\ref{Kraus-representation}) in the case when infidelity is
dominated by decoherence.

    The approximation (\ref{comp-err}) neglects the second-order
corrections due to possibly significant coherent errors. While this
is good enough for the off-diagonal elements of $\chi^{\rm err}$,
let us use a better approximation for the more important diagonal
elements (except $\chi^{\rm err}_{00}$), explicitly taking into
account the top rows and left columns of $\chi^{\rm err}_1$ and
$\chi^{\rm err}_2$ in Eq.\ (\ref{comp-gen}),
    \be
   \chi_{nn}^{\rm err} \approx F_2   \chi_{1,nn}^{\rm err}+
    F_1 \chi_{2,nn}^{\rm err} + 2\, {\rm Im}(\chi^{\rm err}_{1, 0n})\,
    {\rm Im}(\chi^{\rm err}_{2, 0n}),
    \label{comp-diag}\ee
    where the formally similar term $2\, {\rm Re}(\chi^{\rm err}_{1, 0n})\,
    {\rm Re}(\chi^{\rm err}_{2, 0n})$ is neglected, because the elements ${\rm
Re}(\chi_{0n}^{\rm err})$ cannot be relatively large, in contrast to
${\rm Im}(\chi_{0n}^{\rm err})$.

   For the fidelity the exact and approximate
results are
    \begin{eqnarray}
&&  F_\chi =\chi_{00}^{\rm err}=
 \sum_{m,n} \chi_{1,mn}^{\rm err} \chi_{2,mn}^{\rm err}
 \label{comp-err-fid1} \\
&& \hspace{0.5cm} \approx F_1 F_2 - 2\sum\nolimits_{n\neq 0} {\rm
Im} (\chi^{\rm err}_{1,0n})\, {\rm Im} (\chi^{\rm err}_{2,0n}) ,
\qquad
    \label{comp-err-fid2}\end{eqnarray}
 where the similar term $- 2\sum\nolimits_{n\neq 0}
{\rm Re} (\chi^{\rm err}_{1,0n})\, {\rm Re} (\chi^{\rm err}_{2,0n})$
is neglected again.
   Note that ${\rm Tr}(\chi^{\rm err})$ calculated
using approximations (\ref{comp-diag}) and (\ref{comp-err-fid2}) is
slightly smaller than 1, but the difference is of second order in
infidelity.

    The reason for taking a special care of the elements
${\rm Im}(\chi^{\rm err}_{1,0n})$ and ${\rm Im}(\chi^{\rm
err}_{2,0n})$ becomes clear if we consider the composition of small
unitaries $U^{\rm err}_1$ and $U^{\rm err}_2$. Then the approximate
addition (\ref{comp-err-2}) is valid for the first-order
off-diagonal elements of $\chi^{\rm err}$ (which are in the left
column and top row); however, the diagonal elements are of second
order, and for them the errors add up ``coherently'', generating the
``interference'' term $2\, {\rm Im}(\chi^{\rm err}_{1, 0n})\, {\rm
Im}(\chi^{\rm err}_{2, 0n})$ in Eq.\ (\ref{comp-diag}).
    Note that the neglected terms
$2\, {\rm Re}(\chi^{\rm err}_{1, 0n})\,  {\rm Re}(\chi^{\rm err}_{2,
0n})$ have a somewhat similar origin, considering the ``coherent''
composition of the main (``no-jump'') Kraus operators in the
representation (\ref{Kraus-representation}).

     We emphasize that Eqs.\ (\ref{comp-err})--(\ref{comp-err-fid2})
do not change if we exchange the sequence of $\chi^{\rm err}_1$ and
$\chi^{\rm err}_2$. So in this approximation small imperfections of
quantum ``memory'' operations commute with each other, as
intuitively expected. Note that for the fidelity
(\ref{comp-err-fid1}) this result is exact: commutation of arbitrary
error processes does not change $F_\chi$.

\vspace{0.3cm}

    So far we assumed $U_1=U_2=\openone$. Now let us consider
arbitrary desired unitaries  $U_1$ and $U_2$. Then $\chi^{\rm err}$
for the composition can be calculated in two steps (see Fig.\ 2): we
first exchange the sequence of $\chi^{\rm err}_1$ and $U_2$, thus
producing the effective error process $\chi^{\rm err}_{1(U_2)}$, and
then use the discussed above rule for the composition of two
``memory'' operations $\chi^{\rm err}_{1(U_2)}$ and $\chi^{\rm
err}_2$.
    The transformation of $\chi^{\rm err}_1$ when it ``jumps over''
$U_2$ (Fig.\ 2) is essentially the same as the transformation
between $\tilde\chi^{\rm err}$ and $\chi^{\rm err}$ [see Fig.\ 1]
and is given by the equation
    \be
\chi^{\rm err}_{1(U_2)} = W_{(U_2)} \chi^{\rm
err}_{1}W_{(U_2)}^\dagger,
    \label{chi-err-jump}\ee
where $W_{(U_2)}$ is given by Eq.\ (\ref{W-def}).
  This transformation can also be written as
    \be
\chi^{\rm err}_{1(U_2)} = \chi^{\bf I}+
 W_{(U_2)} (\chi^{\rm err}_{1}-\chi^{\bf I}) W^\dagger_{(U_2)} ,
    \ee
so that only the small difference from the ideal memory operation
$\chi^{\bf I}$ is being transformed. Note that this transformation
does not change fidelity, $\chi^{\rm err}_{1(U_2),00}=\chi^{\rm
err}_{1,00}=F_1$.

    Thus we have a relatively simple procedure to find
$\chi^{\rm err}$ for the composition of two quantum operations: we
first apply the transformation (\ref{chi-err-jump}) to move the two
error processes together (Fig.\ 2) and then apply approximate rules
(\ref{comp-err})--(\ref{comp-err-fid2}) to the matrices $\chi^{\rm
err}_{1(U_2)}$ and $\chi^{\rm err}_2$.

\vspace{0.3cm}

    The similar procedure can be used to calculate
$\tilde\chi^{\rm err}$ for the composition of two quantum
operations. We should first move $\tilde\chi^{\rm err}_2$ to the
left by jumping it over $U_1$,
    \be
    \tilde\chi^{\rm err}_{2(U_1)}= W^\dagger_{(U_1)}  \tilde\chi^{\rm
    err}_{2}W_{(U_1)} ,
    \label{chi-2-U1}\ee
and then use the approximate composition rules
(\ref{comp-err})--(\ref{comp-err-fid2}) for the error matrices
$\chi^{\rm err}_{1}$ and $\chi^{\rm err}_2(U_1)$.

For the composition of several quantum operations $\chi^{\rm
err}_i$, the error processes should be first moved to the end of the
sequence by jumping them over the desired unitaries (or moved to the
beginning if we consider the language of $\tilde\chi^{\rm err}$) and
then we use the composition rules
(\ref{comp-err})--(\ref{comp-err-fid2}). The procedure further
simplifies if we can neglect ``coherent'' errors and use the simple
additive rule (\ref{comp-err-2}) for all elements (except $\chi^{\rm
err}_{00}=1-\sum_{n\neq 0}\chi^{\rm err}_{nn}$).

    \section{Unitary corrections}
    \label{sec-corr}

    In experiments it is often useful to check how large
the inaccuracy is of the unitary part of a realized quantum gate and
find the necessary unitary corrections to improve fidelity of the
gate. The error matrix $\chi^{\rm err}$ (or $\tilde\chi^{\rm err}$)
gives us a simple way to do this, because small unitary
imperfections directly show up as the imaginary parts of the
elements $\chi^{\rm err}_{0n}$ and $\chi^{\rm err}_{n0}$.

    Let us assume that we apply a small unitary correction
$U^{\rm corr}=\sum_n u^{\rm corr}_n E_n \approx \openone$ after an
operation characterized by the desired $U$ and error matrix
$\chi^{\rm err}$ with fidelity $F_\chi$. Then the process matrix
$\chi^{\rm corr}_{mn}=u_m^{\rm corr}u_n^{{\rm corr}*}$ for the
correction operation mainly consists of the element $\chi^{\rm
corr}_{00}\approx 1$ and the imaginary first-order elements
$\chi_{n0}^{\rm corr}\approx \chi_{0n}^{{\rm corr}*}\approx u^{\rm
corr}_n$ [see Eq.\ (\ref{imag-unitary})], while all other elements
are of second order. Using Eq.\ (\ref{comp-diag}) we see that after
the correction the elements in the left column of the error matrix
approximately change as
    \be
\chi_{n0}^{\rm err} \rightarrow \chi_{n0}^{\rm err} + F_\chi
u_n^{\rm corr},
    \label{corr-change}\ee
so to correct the unitary imperfection we need to choose
    \be
u_n^{\rm corr} \approx - i {\rm Im} (\chi_{n0}^{\rm err})/F_\chi,
    \label{corr-unitary}\ee
which cancels the imaginary part of the left-column elements (here
the factor $F_\chi^{-1}$ needs an implicit assumption that the
infidelity is dominated by decoherence). The increase of the gate
fidelity $\Delta F_{\chi}$ due to this correction procedure can be
estimated using Eq.\ (\ref{comp-err-fid2}),
    \be
    \Delta F_{\chi}\approx \sum\nolimits_{n\neq 0} ({\rm Im}\,\chi^{\rm
    err}_{n0})^2/F_\chi .
    \label{Delta-F}\ee
In this derivation the factor of 2 in the second term of Eq.\
(\ref{comp-err-fid2}) is compensated by the fidelity decrease due to
the first term. The result (\ref{Delta-F}) in the case
$F_\chi\approx 1$ coincides with what we would expect from the
unitary correction in absence of decoherence. The factor
$F_\chi^{-1}$ in Eq.\ (\ref{Delta-F}) implicitly assumes that the
infidelity is dominated by decoherence.

 Note that the fidelity increase $\Delta F_{\chi}$ is of
second order, so in an experiment we should not expect a significant
improvement of fidelity due to unitary correction, unless the
unitary imperfection is quite big. Also note that in an experiment
it may be easy to apply a unitary correction only in some
``directions'', for example, by applying single-qubit pulses, while
other corrections may be very difficult. In this case only some of
the elements ${\rm Im}(\chi^{\rm err}_{n0})$ can be compensated, and
then the fidelity improvement is given  by Eq.\ (\ref{Delta-F}), in
which summation is only over the elements, compensated by the
correction procedure.

    The above analysis of the compensation procedure assumes a small
compensation. If the unitary error is large, then to find the
optimal correction $U^{\rm corr}$ we can use an iterative procedure,
in which we first estimate the correction via Eq.\
(\ref{corr-unitary}), then use the exact composition relation
(\ref{comp-gen}), and then again adjust the correction via Eq.\
(\ref{corr-unitary}).

    Our analysis of the unitary compensation procedure assumed
application of $U^{\rm corr}$ after the quantum gate. If the
compensation is applied before the quantum gate, then it is more
natural to use the language of $\tilde\chi^{\rm err}$ (see Fig.\ 1);
in this case Eqs.\ (\ref{corr-change})--(\ref{Delta-F}) remain
valid, with $\chi^{\rm err}$ replaced by $\tilde\chi^{\rm err}$.
Applying corrections both before and after the gate may in some
cases increase the number of correctable ``directions'' in the space
of unitary operators.

\vspace{0.3cm}

     To illustrate analysis of the unitary corrections, let us
consider the two-qubit controlled-Z (CZ) gate in the ``quantum von
Neumann architecture''
\cite{Matteo-Science,Lucero-12,Galiautdinov-12}, in which
single-qubit $Z$-rotations are realizable very easily (without an
additional cost), and so such corrections can be easily applied.
Application of $Z$-rotation over the small angle $\varphi_1$ to the
first qubit and $Z$-rotation over the small angle $\varphi_2$ to the
second qubit produces the correction unitary $U^{\rm corr}={\rm
diag}(1,e^{i\varphi_1}, e^{i\varphi_2}, e^{i\varphi_3})$ in the
basis $\{ |00\rangle, |10\rangle, |01\rangle,|11\rangle \}$. Here
$\varphi_3=\varphi_1+\varphi_2$, but if we can also introduce
correction $\varphi_{CZ}$ of the CZ angle, then it will be
$\varphi_3=\varphi_1+\varphi_2+\varphi_{CZ}$.
   Expansion of $U^{\rm corr}$ in the Pauli basis gives four
non-zero elements:
    \begin{eqnarray}
&&    u^{\rm corr}_{II}= \frac{1}{d}{\rm Tr}(U^{\rm corr} \times
II)=
    \frac{1}{4} (1+e^{i\varphi_1}+e^{i\varphi_2}+e^{i\varphi_3})
    \nonumber \\
&&  \hspace{1cm}   \approx 1+i(\varphi_1+\varphi_2+\varphi_3)/4,
    \label{u-corr-II}\\
&& u^{\rm corr}_{IZ}= \frac{1}{d}{\rm Tr}(U^{\rm corr} \times IZ) =
\frac{1}{4} (1-e^{i\varphi_1}+e^{i\varphi_2}-e^{i\varphi_3})
    \nonumber \\
&& \hspace{1cm}  \approx i(-\varphi_1+\varphi_2-\varphi_3)/4, \qquad
     \label{u-corr-IZ}\\
&& u^{\rm corr}_{ZI}= \frac{1}{d}{\rm Tr}(U^{\rm corr} \times ZI)
\approx \frac{i}{4}(\varphi_1-\varphi_2-\varphi_3), \qquad
     \label{u-corr-ZI}\\
&& u^{\rm corr}_{ZZ}= \frac{1}{d}{\rm Tr}(U^{\rm corr} \times ZZ)
\approx \frac{i}{4}(-\varphi_1-\varphi_2+\varphi_3), \qquad
     \label{u-corr-ZZ}\end{eqnarray}
where we used the standard notation for the two-qubit operator basis
$\{ II, IX, IY, IZ, XI, \dots ZZ \}$; note that the index 0 in the
notations of our paper corresponds to $II$.

    It is important to emphasize that this expansion of
$U^{\rm corr}$ is not exactly what we used in Eqs.\
(\ref{corr-change}) and (\ref{corr-unitary}) because  we assumed
real $u^{\rm corr}_0\approx 1$, while $u_{II}^{\rm corr}$ in Eq.\
(\ref{u-corr-II}) is not real. We therefore need to adjust the
overall phase of $U^{\rm corr}$ to make $u_{II}^{\rm corr}$ real.
This can be easily done by replacing $U^{\rm corr}$ with  $U^{\rm
corr} e^{-i(\varphi_1+\varphi_2+\varphi_3)/4}$. However, for small
angles $\varphi_i$ this would produce only a small change in Eqs.\
(\ref{u-corr-IZ})--(\ref{u-corr-ZZ}). The left column of $\chi^{\rm
corr}$ therefore contains the same non-zero elements,
    \begin{eqnarray}
  && \chi^{\rm corr}_{IZ,II}  \approx i(-\varphi_1+\varphi_2-\varphi_3)/4
  =i(-2\varphi_1-\varphi_{CZ})/4, \qquad
     \\
&& \chi^{\rm corr}_{ZI,II} \approx i
(\varphi_1-\varphi_2-\varphi_3)/4=i (-2\varphi_2-\varphi_{CZ})/4 ,
\qquad
     \\
&& \chi^{\rm corr}_{ZZ,II} \approx
i(-\varphi_1-\varphi_2+\varphi_3)/4=i\varphi_{CZ}/4. \qquad
    \end{eqnarray}

    If we correct only $\varphi_1$ and $\varphi_2$
(so that $\varphi_{CZ}=0$), then we should choose them [see Eq.\
(\ref{corr-unitary})] as
    \be
    \varphi_1 \approx 2\, {\rm Im} (\chi^{\rm err}_{IZ,II})/F_\chi ,
    \,\,\,
 \varphi_2 \approx 2\, {\rm Im} (\chi^{\rm err}_{ZI,II})/F_\chi,
    \label{correct-phi12}\ee
where $\chi^{\rm err}$ and $F_\chi=\chi^{\rm err}_{II,II}$ are
measured experimentally. This will cancel the left-column elements
${\rm Im}(\chi^{\rm err}_{IZ,II})$ and ${\rm Im}(\chi^{\rm
err}_{ZI,II})$ in the corrected quantum gate and produce fidelity
improvement $\Delta F_\chi \approx [({\rm Im} \chi^{\rm
err}_{IZ,II})^2+ ({\rm Im} \chi^{\rm err}_{ZI,II})^2]/F_\chi $.

    If we can also correct $\varphi_{CZ}$, then we should choose
corrections
    \begin{eqnarray}
&&       \varphi_1 \approx 2\, {\rm Im} (\chi^{\rm err}_{IZ,II}+
\chi^{\rm err}_{ZZ,II})/F_{\chi} ,
    \\
&& \varphi_2 \approx 2\, {\rm Im} (\chi^{\rm err}_{ZI,II}
 + \chi^{\rm err}_{ZZ,II})/F_\chi,
   \\
&&   \varphi_{CZ}\approx - 4 \, {\rm Im} (\chi^{\rm
err}_{ZZ,II})/F_\chi .
    \label{correct-CZphase}\end{eqnarray}
This will cancel the left-column elements ${\rm Im}(\chi^{\rm
err}_{IZ,II})$, ${\rm Im}(\chi^{\rm err}_{ZI,II})$, and ${\rm
Im}(\chi^{\rm err}_{ZZ,II})$  in the corrected quantum gate and
produce fidelity improvement $\Delta F_\chi \approx [({\rm Im}
\chi^{\rm err}_{IZ,II})^2+ ({\rm Im} \chi^{\rm err}_{ZI,II})^2+
({\rm Im} \chi^{\rm err}_{ZZ,II})^2]/F_\chi $.

    Note that $U^{\rm corr}$ commutes with the CZ gate,
$U^{CZ}={\rm diag}(1,1,1,-1)$, so it does not matter if the
single-qubit corrections are applied before or after the gate (the
$\varphi_{CZ}$ correction is obviously a correction of the gate
itself). Similarly, it does not matter if we use $\chi^{\rm err}$ or
$\tilde\chi^{\rm err}$ in this correction procedure.

\section{Error matrix from the Lindblad-form decoherence}
\label{sec-Lindblad-error}

    Let us consider a quantum evolution described by the Lindblad-form
master equation
    \be
    \dot\rho  = -\frac{i}{\hbar} [H,\rho] +\sum_j \Gamma_j
    (B_j \rho B_j^\dagger - \frac{1}{2}B^\dagger_jB_j \rho
    -\frac{1}{2}\rho B^\dagger_jB_j) ,
    \label{Lindblad-evol}\ee
where $H(t)$ is the Hamiltonian (which has a significant time
dependence for a multi-stage quantum gate) and $j$th decoherence
mechanism is described by the Kraus operator $B_j$ and the rate
$\Gamma_j(t)$. Mathematically it is natural to work with the
combination $\sqrt{\Gamma_j}\, B_j$; however, we prefer to keep
$\Gamma_j$ and $B_j$ separate because they both have clear physical
meanings (see Appendix B).

    The imperfection of a quantum gate comes from imperfect control
of the Hamiltonian $H(t)$ and from decoherence. If both
imperfections are small, we can consider them separately. So, in
this section we assume perfect $H(t)$ and analyze the process error
matrix due to decoherence only. Moreover, we will consider only one
decoherence mechanism, since summation over them is simple and can
be done later. Therefore we will drop the index $j$ in Eq.\
(\ref{Lindblad-evol}) and characterize the decoherence process by
$B$ and $\Gamma (t)$.

    To find the error matrix $\chi^{\rm err}$ (or
$\tilde\chi^{\rm err}$) of such operation, we can divide the total
gate duration $t_G$ into small timesteps $\Delta t$, for each of
them representing the evolution as the desired unitary
$\exp[-iH(t)\,\Delta t]$ and the error process $\chi^{\rm
err}(t,\Delta t)$. Then using the same idea as in section
\ref{sec-comp}, we can jump the error processes over the unitaries,
moving them to the very end (for $\chi^{\rm err}$) or to the very
beginning of the gate (for $\tilde\chi^{\rm err}$). Finally, we can
add up the error processes for all $\Delta t$ using approximations
(\ref{comp-err})--(\ref{comp-err-fid2}).

    For small $\Delta t$ the error matrix
$\chi^{\rm err}(t,\Delta t)$ can be found by expanding $B$ and
$B^\dagger B$ in the Pauli basis, $B=\sum_n b_n E_n$, $B^\dagger
B=\sum_n c_n E_n$, and then comparing the decoherence terms in Eq.\
(\ref{Lindblad-evol}) with Eq.\ (\ref{chi-def}),
    \begin{eqnarray}
    &&  \chi^{\rm err}_{mn}(t,\Delta t)= \chi^{\bf I}_{mn}
    +\Gamma \Delta t \, {\cal B}_{mn},
    \label{chi-err-B-mn}\\
    && {\cal B}_{mn}
    = b_m b_n^* - \frac{1}{2} (c_m\delta_{n0}
+c_n^*\delta_{m0}) , \qquad
    \label{B-mn}\\
&& b_n =\frac{1}{d} \, {\rm Tr} (B E_n^\dagger), \,\,\,
    \\
&& c_n=\frac{1}{d} \, {\rm Tr}(B^\dagger B E_n^\dagger)= \frac{1}{d}
\, \sum_{p,q} b_p^*b_q {\rm Tr} (E_p^\dagger E_q E_n^\dagger ).
\qquad
    \label{c-n}\end{eqnarray}
It is easy to see that $c_n=c_n^*$ since $E_n^\dagger=E_n$ for the
Pauli basis, so the left-column and top-row contributions due to the
$c$-terms in Eq.\ (\ref{B-mn}) are real. Note that we can always use
the transformation $B\rightarrow B- b_0 \openone$ in the Lindblad
equation, which compensates (zeroes) the component $b_0$ but changes
the Hamiltonian, $H\rightarrow H+H_a$ with $H_a=i\hbar (\Gamma/2)
(b_0^*B-b_0 B^\dagger)$ (there is no change, $H_a=0$, if $B$ is
Hermitian). Therefore we can use $b_0=0$ in Eq.\ (\ref{B-mn}), and
then the left-column and top-row elements come only from the
$c$-terms, which correspond to the the terms $-(B^\dagger B\rho
-\rho B^\dagger B)/2$ of the Lindblad equation and so correspond to
the ``no-jump'' evolution (see Appendix B).

    Thus we have a clear physical picture of where the components
of $\chi^{\rm err}(t,\Delta t)-\chi^{\bf I}$ come from: the
imaginary parts of the left-column and top-row elements come from
the unitary imperfection (which may also be related to the
decoherence-induced change of the Hamiltonian), the real parts of
the left-column and top-row elements come from the ``no-jump''
evolution (see Appendix B), and other elements come from the
decoherence ``jumps'', which ``strongly'' change the state (recall
that $B$ have only components orthogonal to $\openone$). A similar
interpretation has been used in Sec.\ \ref{sec-properties}.
 Note that $\tilde\chi^{\rm
err}(t,\Delta t)=\chi^{\rm err}(t,\Delta)$ for small $\Delta t$
because there is practically no unitary evolution.

    Now let us use the language of $\tilde\chi^{\rm err}$, which
relates the error process to the beginning of the gate ($t=0$). We
can find $\tilde\chi^{\rm err}$ by moving the error processes
$\tilde\chi^{\rm err}(t,\Delta t)$ to the start of the gate using
the transformation relation (\ref{chi-2-U1}) and then summing up the
error contributions  using the approximate additive rule
    (\ref{comp-err-matrix}). In this way we obtain
    \begin{eqnarray}
&& \tilde\chi^{\rm err} \approx \chi^{\bf I} +\int_0^{t_G} \Gamma \,
W^\dagger (t)\,
 {\cal B}  \, W (t) \, dt,
     \label{Lindblad-chi-err-tilde}\\
&&  W_{mn}(t)= \frac{1}{d} \, {\rm Tr} [E_m^\dagger U (t) E_n
U^\dagger (t)],
    \\
    &&  U(t)=\exp [\frac{-i}{\hbar }\int_0^t H(t)\, dt],
    \label{U(t)}  \end{eqnarray}
where ${\cal B}$ is given by Eq.\ (\ref{B-mn}), $U(t)$ is the
unitary evolution occurring within the interval $(0,t)$, and Eq.\
(\ref{U(t)}) assumes the time-ordering of operators.

    Similar procedure can be used to find $\chi^{\rm err}$; then we
should move the errors to the end of the gate, by jumping them over
the remaining unitary $U_{\rm rem}(t)=U(t_G) U^\dagger (t)$,
    \begin{eqnarray}
&& \chi^{\rm err} \approx \chi^{\bf I} +\int_0^{t_G} \Gamma \,
W_{\rm rem}(t)\,
 {\cal B}  \, W_{\rm rem}^\dagger (t) \, dt,
    \label{Lindblad-chi-err}\\
&&  W_{{\rm rem},mn}(t)= \frac{1}{d}\, {\rm Tr} [E_m^\dagger U(t_G)
U^\dagger (t) E_n U(t) U^\dagger(t_G)] . \qquad
     \label{Lindblad-W}\end{eqnarray}

Note that the $W_{\rm rem}$-transformation of ${\cal B}$ in Eq.\
(\ref{Lindblad-chi-err}), ${\cal B}(t)\equiv W_{\rm rem}(t)\,
 {\cal B}  \, W_{\rm rem}^\dagger (t)$, is equivalent to the
 $U_{\rm rem}$-transformation of the Kraus operator $B$, which
 relates it to the end of the gate,
    \be
B(t)=U_{\rm rem}(t) B U_{\rm rem}^ \dagger(t).
    \label{B(t)}\ee
  Therefore instead of
Eq.\ (\ref{Lindblad-chi-err}) we can use
    \be
\chi^{\rm err}\approx \chi^{\bf I}+\int \Gamma {\cal B}(t)\, dt,
    \label{Lindblad-chi-err-2}\ee
in which ${\cal B}(t)$ is given by Eqs.\ (\ref{B-mn})--(\ref{c-n})
using $B(t)$ instead of $B$.
   Similarly, instead of Eq.\ (\ref{Lindblad-chi-err-tilde}) we can
use
    \be
\tilde \chi^{\rm err}\approx \chi^{\bf I}+\int \Gamma \tilde{\cal
B}(t)\, dt,
    \ee
in which $\tilde{\cal B}(t)$ is given by Eqs.\
(\ref{B-mn})--(\ref{c-n}) with
    \be
\tilde{B}(t)\equiv U^\dagger(t) B U(t)
    \label{tilde-B}\ee
 instead of $B$.
    Note that if $B$ is orthogonal to $\openone$ ($b_0=0$,
see discussion above), then $B(t)$ and $\tilde{B}(t)$ are also
orthogonal to $\openone$, so they still describe ``strong'' error
jumps.

    Since the element ${\cal B}_{00}=|b_0|^2-c_0=-\sum_{n\neq0} |b_n|^2$
does not change in the transformation $W_{\rm rem}{\cal B}W_{\rm
rem}^\dagger$, in the approximation (\ref{Lindblad-chi-err}) the
fidelity is
    \be
    F_{\chi} \approx 1-\int_0^{t_G} \Gamma \sum_{n\neq 0} |b_n|^2 \,
    dt,
    \label{Lindblad-F}\ee
so that if $\Gamma$ does not depend on time, then $F_\chi \approx
1-t_G \Gamma  \sum_{n\neq 0} |b_n|^2$ decays linearly with the gate
time $t_G$.
 (As discussed above, the element $b_0$ is equivalent to a unitary
imperfection, and therefore brings infidelity, which scales
quadratically with time.)
 An interesting observation is
that in this approximation the fidelity does not depend on the
desired unitary evolution $U(t)$.
 Therefore, for example, for {\it any two-qubit gate}
(which does not involve higher physical levels in the qubits) the
contribution to the infidelity due to the energy relaxation and
(Markovian non-correlated) pure dephasing of the qubits is
    \be
    1-F_\chi = \frac{t_G}{2T_1^{(a)}}+\frac{t_G}{2T_1^{(b)}}
    + \frac{t_G}{2T_\varphi^{(a)}} + \frac{t_G}{2T_\varphi^{(b)}},
    \label{Lindbad-T1-T2}\ee
where $T_1$ is the energy relaxation time, $T_\varphi$ is the pure
dephasing time, and the qubits are labeled by superscripts $(a)$ and
$(b)$ (see Appendix A). The independence of the fidelity
(\ref{Lindblad-F}) on the unitary evolution can be understood in the
following way. The process fidelity $F_\chi$ is related to the state
fidelity, uniformly averaged over all pure initial states. A unitary
evolution does not change the uniform distribution of pure states;
therefore, the average rate of rare ``decoherence jumps'' does not
depend on the unitary part.

    The approximation (\ref{Lindblad-chi-err-tilde})--(\ref{Lindblad-F})
neglects the second-order corrections.  In the case of significant
coherent errors (which is not typical when only Lindblad-equation
decoherence is discussed), the natural second-order correction is to
add $\chi_{m0}^{\rm err} (\chi_{0n}^{\rm err})^*$ to the elements
$\chi^{\rm err}_{mn}$ with $m\neq 0$ and $n\neq 0$ (we assume
$1-F_\chi\ll 1$). Since this correction is small, it may be
important only for the diagonal elements $\chi^{\rm err}_{nn}$,
because it affects the resulting fidelity $F_\chi=1-\sum_{n\neq 0}
\chi^{\rm err}_{nn}$. This will introduce correction into Eq.\
(\ref{Lindblad-F}), which scales quadratically with the gate time
$t_G$. The similar second-order correction  $\tilde\chi_{m0}^{\rm
err} (\tilde\chi_{0n}^{\rm err})^*$ can be introduced to the
elements of $\tilde\chi^{\rm err}_{mn}$.

    Note that in Eqs.\ (\ref{Lindblad-chi-err}) and
(\ref{Lindblad-chi-err-2}) the elements of $\chi^{\rm err}$ are
linear in the decoherence rate $\Gamma$.
 Therefore the
``pattern'' of $\chi^{\rm err}$ elements is determined by the
decoherence mechanism characterized by the operator ${\cal B}$ and
its transformation $W_{\rm rem}(t)$, and this pattern is multiplied
by the decoherence rate $\Gamma$. (The experimental  $\chi^{\rm
err}$ may need subtraction of the discussed above second-order
correction to become linear in $\Gamma$.)

    If there are several decoherence mechanisms in Eq.\
(\ref{Lindblad-evol}), then in the first order their contributions
to $\chi^{\rm err}$ simply add up. Therefore, if the patterns for
the different decoherence mechanisms $B_j$ are sufficiently simple
and distinguishable from each other, then the decoherence rates
$\Gamma_j$ can be found directly from the experimentally measured
$\chi^{\rm err}$ (again, subtraction of the second-order correction
may be useful in the case of significant coherent errors).

    \section{SPAM identification}
    \label{sec-SPAM}

    A very important difficulty in experimental implementation of the
QPT is due to SPAM errors: imperfect preparation of the initial
states and state tomography errors, which include imperfect
tomographic single-qubit rotations and imperfect measurement of
qubits. In this section we discuss a way, which may help solving
this problem.

    First, let us assume that the imperfect state preparation can be
represented as an error channel, which acts on the ideal initial
state. If we use $4^N$ initial states of $N$ qubits, then the
transformation between $4^N$ ideal and real density matrices of the
initial states can always be described by a linear transformation,
characterized by $16^N$ parameters. So by the number of parameters
it seems that the representation of the preparation error by an
error channel is always possible. The problem, however, is that this
transformation may happen to be non-positive. Also, if more than
$4^N$ initial states are used in an experiment, then an
error-channel representation may be impossible by the number of
parameters. Nevertheless, we will use this representation, arguing
that it can somehow be introduced phenomenologically.
    Similarly, we assume that the imperfections of the tomographic
single-qubit rotations and measurement can also be represented as an
error channel.

\begin{figure}[tb]
  \centering
\includegraphics[width=8.0cm]{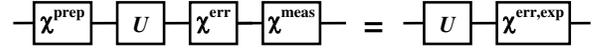}
 \vspace{-0.2cm}
  \caption{Representation of the SPAM errors by the error processes
$\chi^{\rm prep}$ and  $\chi^{\rm prep}$. }
  \label{fig-SPAM}
\end{figure}

    Using these two assumptions, we describe the preparation errors by
the error matrix $\chi^{\rm prep}$ (which is close to the ideal
memory $\chi^{\bf I}$), and the tomography/measurement errors are
described by $\chi^{\rm meas}$ (also close to $\chi^{\bf I}$) -- see
Fig.\ 3. Thus the experimentally measured error matrix $\chi^{\rm
err,exp}$ is due to $\chi^{\rm prep}$, $\chi^{\rm err}$ (which we
need to find) and $\chi^{\rm meas}$.

   The general idea is to measure $\chi^{\rm prep}$ and
$\chi^{\rm meas}$ by doing the process tomography without the gate
(doing the tomography immediately on the initial states) and then
subtract this SPAM error from $\chi^{\rm err,exp}$ to obtain
$\chi^{\rm err}$. So, the procedure which first comes to mind is to
use $\chi^{\rm err}\approx \chi^{\rm err,exp}- (\chi^{{\rm
err,}I}-\chi^{\bf I})$, where $\chi^{{\rm err,}I}$ is the
experimentally measured $\chi$ without the gate. However, in general
this would be wrong. The reason is that without the gate we measure
the simple sum of the SPAM-error components [see Eq.\
(\ref{comp-err-matrix})],
    \be
\chi^{{\rm err,}I}\approx \chi^{\rm prep}+ \chi^{\rm meas}
-\chi^{\bf I},
    \ee
but in the presence of the gate $U$ the preparation error $\chi^{\rm
prep}$ changes because it is ``jumped over'' $U$ [see Sec.\
\ref{sec-comp} and Eq.\ (\ref{chi-err-jump})], so that
    \be
    \chi^{\rm err,exp} \approx \chi^{\rm err} + W_{(U)}(\chi^{\rm prep}
    -\chi^{\bf I})W_{(U)}^\dagger + (\chi^{\rm meas}  -\chi^{\bf I}),
    \label{spam-exp}\ee
where $W_{(U)}$ is given by Eq.\ (\ref{W-def}). Therefore to find
the actual error matrix $\chi^{\rm err}$ from the experimental
$\chi^{\rm err,exp}$, it is insufficient to know $\chi^{{\rm
err,}I}$. We can still find  $\chi^{\rm err}$  from Eq.\
(\ref{spam-exp}), but we need to know $\chi^{\rm prep}$ and
$\chi^{\rm meas}$ separately.

    The idea how to find both $\chi^{\rm prep}$ and
$\chi^{\rm meas}$ is to do a calibration QPT with a set of gates
with very high fidelity, for which the error is negligible, and then
use Eq.\ (\ref{spam-exp}) to separate the changing contribution from
$\chi^{\rm prep}$ and non-changing contribution from $\chi^{\rm
meas}$. For example, in experiments with superconducting qubits the
one-qubit $\pi$ and $\pi/2$ rotations about $X$ and $Y$ axes usually
have much better fidelity than two-qubit or multi-qubit gates.
Therefore, for the QPT of a two-qubit or multi-qubit gate we can
rely on the SPAM-error identification using these one-qubit gates.

    Let us start with considering a single qubit and discussing the change
$\chi^{\rm prep}\rightarrow \chi^{\rm prep}_{(U)}=
 W_{(U)}\chi^{\rm prep}W_{(U)}^\dagger$ of the preparation error
contribution, when we apply a high-fidelity $X$ gate, $U=X$. By
using Eq.\ (\ref{spam-exp}) or by simply comparing the terms in the
equation $\sum_{mn}\chi^{\rm prep}_{(X),mn}E_m X \rho_{\rm in} X
E_n=\sum_{mn}\chi^{\rm prep}_{mn} X E_m\rho_{\rm in} E_n X$, it is
easy to find that this transformation flips the signs of the
off-diagonal elements $IY$, $IZ$, $XY$, and $XZ$  of $\chi^{\rm
prep}$ (the same for the symmetric, complex-conjugate elements),
while the off-diagonal elements $IX$ and $YZ$ (and symmetric
elements) do not change. The diagonal elements do not change, as
expected for a Pauli twirling. Similarly, if we apply the
high-fidelity $Y$ gate, then the off-diagonal elements $IX$, $IZ$,
$XY$, and $YZ$ flip the sign, while the elements $IY$ and $XZ$ do
not change. Recall that the contribution from $\chi^{\rm meas}$ does
not change. Therefore, by comparing $\chi^{\rm err,exp}$ for the
gates $X$, $Y$, and $I$ (no gate), we can find separately all the
off-diagonal elements of $\chi^{\rm prep}$ and $\chi^{\rm meas}$.

    To find diagonal elements of $\chi^{\rm prep}$ and
$\chi^{\rm meas}$, we can use high-fidelity gates $\sqrt{X}$
($\pi/2$ rotation over $X$ axis) and $\sqrt{Y}$ ($\pi/2$ rotation
over $Y$ axis). The gate $\sqrt{X}$ exchanges diagonal elements $YY$
and $ZZ$ of $\chi^{\rm prep}$ (it also exchanges and flips signs of
some off-diagonal elements, but for simplicity we focus on the
diagonal elements only). Similarly, the operation $\sqrt{Y}$
exchanges the elements $XX$ and $ZZ$ of $\chi^{\rm prep}$. Since the
elements of $\chi^{\rm meas}$ do not change, we can find the
diagonal elements of $\chi^{\rm prep}$ and $\chi^{\rm meas}$
separately. Actually, this cannot be done in the unique way, because
the contributions proportional to $\openone$ in $\chi^{\rm
prep}-\chi^{\bf I}$ and in $\chi^{\rm meas}-\chi^{\bf I}$ are
indistinguishable from each other (these are the
depolarization-channel contributions). However, this non-uniqueness
is not important because any choice gives the same SPAM-error
contribution to $\chi^{\rm err,exp}$.

    In this way, by doing QPT of the gates $X$, $Y$, $\sqrt{X}$,
$\sqrt{Y}$ and $I$ for a single qubit, we can find $\chi^{\rm prep}$
and $\chi^{\rm meas}$ (assuming that these gate are nearly perfect
in comparison with preparation and measurement errors). For two or
more qubits we can do the similar procedure, applying the
combinations of these 5 single-qubit gates, and thus identifying all
the elements of the multi-qubit matrices $\chi^{\rm prep}$ and
$\chi^{\rm meas}$. In fact, the system of equations for this
identification is overdetermined, so we can either use an ad-hoc way
of calculating the elements or use the numerically efficient
least-square method (via the pseudo-inverse). Note that the
described procedure has an obvious relation to the Pauli and
Clifford twirling, but for us it is sufficient to use only a small
subset of operations, and we do not average the result.

    The described procedure of the SPAM-error identification
[which is then subtracted from $\chi^{\rm err,exp}$ using Eq.\
(\ref{spam-exp})] is surely very cumbersome. However, this is at
least some way to deal with the SPAM problem, which does not seem to
have a simple solution. Moreover, there are several ways to make
experimental procedure less cumbersome, which are discussed next.

    The situation is greatly simplified if the SPAM-error is
dominated by only one component: either $\chi^{\rm prep}$ or
$\chi^{\rm meas}$. If $\chi^{\rm prep}$ is negligible, then from
Eq.\ (\ref{spam-exp}) we see that the error matrix of the analyzed
multi-qubit gate $U$ can be estimated as
    \be
\chi^{\rm err}\approx \chi^{\rm err,exp}-(\chi^{{\rm
err,}I}-\chi^{\bf I}),
    \ee
so besides the standard QPT of the $U$ gate we only need the QPT of
no operation ($I$ gate). In the opposite limit when $\chi^{\rm
meas}$ is negligible, it is easier to use the language of
$\tilde{\chi}^{\rm err}$, because we can neglect the change of
$\chi^{\rm meas}$ when it is jumped over $U$ to the left. In this
case
    \be
\tilde{\chi}^{\rm err}\approx \tilde{\chi}^{\rm err,exp}-(\chi^{{\rm
err,}I}-\chi^{\bf I});
    \ee
recall that $\tilde{\chi}^{{\rm err,}I}=\chi^{{\rm err,}I}$.

    When both $\chi^{\rm prep}$ and $\chi^{\rm meas}$ are
significant in the SPAM-error, it is still possible to simplify the
described above procedure by using the idea of compressed-sensing
QPT \cite{Flammia-12,Shabani-11,Rodionov-13}. In contrast to the
usual application of the compressed-sensing idea to the QPT of the
gate $U$, we  can apply it to find $\chi^{\rm prep}$ and $\chi^{\rm
meas}$ by using a small random subset of $5^N$ combinations of
single-qubit gates in the procedure. It is also possible to combine
the random choice of the gates with the random choice of initial
states and measurement directions, thus further reducing the amount
of experimental work. It is important to emphasize that we do not
need to know the matrices $\chi^{\rm prep}$ and $\chi^{\rm meas}$
very precisely, so their compressed-sensing estimate should be
sufficient.

    One more idea, which may be practically useful, is to measure
$\chi^{{\rm err},I}$ and select only few significant peaks in it.
For each of these peaks we identify which contribution to it comes
from $\chi^{\rm prem}$ and from $\chi^{\rm meas}$ by applying just
one or a few single-qubit rotations, which change this particular
peak. It is beneficial to choose the rotations, which affect more
than one significant peak of $\chi^{{\rm err},I}$. In this way a
relatively small number of QPTs is sufficient to find the
significant peaks of $\chi^{\rm prep}$ and $\chi^{\rm meas}$. Then
by using Eq.\ (\ref{spam-exp}) we estimate the SPAM contribution for
the multi-qubit gate $U$ and subtract it from the experimental error
matrix $\chi^{\rm err,exp}$ to find ``actual'' $\chi^{\rm err}$.

    Note that we do not need a complicated procedure to find the
``actual'' process fidelity $F_\chi$ of the gate $U$. If the SPAM
errors can be represented by the error channels (Fig.\ 3) and if
there are no significant coherent SPAM errors, then
    \be
    F_\chi \approx F_\chi^{\rm exp}/F_{\chi}^{I},
    \label{spam-fid}\ee
where $F_\chi^{\rm exp}=\chi^{\rm err,exp}_{00}$ and
$F_{\chi}^{I}=\chi^{{\rm err},I}_{00}$ are the experimentally
measured fidelities for the gate $U$ and no gate, respectively [see
Eq.\ (\ref{comp-err-fid2})].

    \section{Conclusion}
    \label{sec-concl}

    In this paper we have discussed representation of quantum
operations via the error matrices. Instead of characterizing an
operation by the standard process matrix $\chi$, we separate the
desired unitary operation $U$ and the error process, which is placed
either after or before $U$ (Fig.\ 1). This defines two error
matrices: $\chi^{\rm err}$ and $\tilde\chi^{\rm err}$ [Eqs.\
(\ref{chi-err-def}) and (\ref{tilde-chi-err-def})]. We use the
standard Pauli basis $\{E_n\}$ for all process matrices. The error
matrices $\chi^{\rm err}$ and $\tilde\chi^{\rm err}$ are related to
$\chi$ via unitary transformations [Eqs.\ (\ref{chi-err-chi}) and
(\ref{chi-tilde-err-chi})], as well as to each other [Eqs.\
(\ref{W-transf}) and (\ref{W-def})]. Therefore the error matrices
are equivalent to $\chi$; however, they are more convenient to use
than $\chi$. The error matrices have only one large element, which
is located at the top left corner and is equal to the process
fidelity $F_\chi$. Any other non-zero element corresponds to an
imperfection of the quantum gate. Therefore, the bar chart of
$\chi^{\rm err}$ (or $\tilde\chi^{\rm err}$) is a visually
convenient way of representing the imperfections of an experimental
quantum gate.

    The elements of $\chi^{\rm err}$ (or $\tilde\chi^{\rm err}$)
have more intuitive physical meaning related to the operation
imperfection, than the elements of $\chi$ (even though the meaning
of most of the elements is still not as intuitive as we would wish).
It is important that since the error-matrix elements are small for a
high-fidelity gate, the first-order approximation is typically
sufficient.
  The imaginary parts of the elements in the left column and top
row correspond to the unitary imperfection,
    $
    U^{\rm err} \approx \openone + \sum_{n>0} i \, {\rm Im}
    (\chi_{n0}^{\rm err}) E_n/ F_\chi,
    $
where the correction factor $1/F_\chi$ can be taken seriously only
if most of infidelity comes from decoherence. The real parts of the
elements in the left column and top row correspond to the small
non-unitary change of the quantum state in the case when no
``jumps'' due to decoherence occur (this change is due to the
Bayesian update, see Appendix B). It is natural to combine this
non-unitary change with the unitary imperfection into a ``coherent''
(or ``gradual'') state change, which happens in absence of
``jumps''. Finally, other elements of $\chi^{\rm err}_{mn}$, with
$m\neq 0$ and $n\neq 0$, correspond to the strong ``jumps'' of the
quantum state due to decoherence. These jumps are characterized by
Kraus operators practically orthogonal to $\openone$ and therefore
always bring an error (see discussions in Secs.\
\ref{sec-properties} and \ref{sec-Lindblad-error}). The diagonal
elements $\chi_{nn}^{\rm err}$ with $n\neq 0$ (probabilities of
$E_n$-type errors in the Pauli twirling approximation
\cite{Ghosh-12,Geller-13}) have contributions from both the
``coherent'' imperfection and decoherence ``jump'' processes;
however, the ``coherent'' contribution to $\chi_{nn}^{\rm err}$ is
of second order (crudely, $|\chi^{\rm err}_{n0}|^2$  or $|\chi^{\rm
err}_{n0}|^2/F_\chi$), so typically the main contribution is
expected to be from decoherence, unless the unitary part is very
inaccurate. We mainly discuss $\chi^{\rm err}$, but everything is
practically the same in the language of $\tilde\chi^{\rm err}$.

    The composition of two error processes in the absence of desired
unitary operations can be represented in the first order as a simple
addition of the corresponding error matrices [Eqs.\
(\ref{comp-err})--(\ref{comp-err-matrix})]. However, if for $M$
sequential error processes the ``coherent'' elements ${\rm
Im}(\chi_{n0}^{\rm err})$ add up with the same phases and thus the
sum grows linearly with $M$, then the second-order contribution (in
particular, to the diagonal elements $\chi_{nn}^{\rm err}$) grows as
$M^2$, and for large $M$ it can become significant in comparison
with the first-order decoherence contribution, which grows linearly
with $M$. For a composition of two quantum gates with non-trivial
desired unitaries we need first to ``jump'' the error process over
the unitary (see Fig.\ 2), that is described by the transformation
(\ref{chi-err-jump}), and then add the error matrices.

    Essentially the same procedure can be done to calculate the
error matrix contribution due to the Lindblad-form decoherence in a
quantum gate, which has finite duration and non-trivial evolution in
time. For each short time step $\Delta t$ the decoherence produces a
contribution to $\chi^{\rm err}$ [Eq.\ (\ref{chi-err-B-mn})], but
this contribution should be ``jumped over'' the unitary evolution to
the beginning or the end of the gate before being summed up [Eqs.\
(\ref{Lindblad-chi-err-tilde}) and (\ref{Lindblad-chi-err})]. The
equivalent language is to ``jump'' the decoherence Kraus operators
over the desired unitaries, before the summation of error matrices
[Eqs.\ (\ref{B(t)})--(\ref{tilde-B})]. It is interesting that in the
leading order the contribution to the infidelity $1-F_\chi$ from the
decoherence (if it occurs within the same Hilbert space) does not
depend on the desired unitary evolution [Eqs.\ (\ref{Lindblad-F})
and (\ref{Lindbad-T1-T2})].

    Since the elements ${\rm Im}(\chi^{\rm err}_{n0})$ directly tell us about
the unitary imperfection, it is easy to find the needed unitary
correction [Eq.\ (\ref{corr-unitary})] and the corresponding
fidelity improvement [Eq.\ (\ref{Delta-F})]. However, the fidelity
improvement is of second order and therefore is typically not
expected to be significant. We have considered a particular example
of correcting a CZ gate using single-qubit $Z$-rotations and
CZ-phase corrections [Eqs.\
(\ref{correct-phi12})--(\ref{correct-CZphase})].

    The QPT suffers from errors in preparing the initial states
and tomography measurement (SPAM errors). While this problem does
not seem to have a simple solution, in Sec.\ \ref{sec-SPAM} we have
discussed a way, which may be helpful in alleviating this problem. A
natural idea is to measure the error matrix $\chi^{{\rm err},I}$ in
the absence of the gate and then subtract it from the measured error
matrix $\chi^{\rm err,exp}$ of the characterized gate $U$. However,
this idea works only if the SPAM is dominated by one type of error:
either at the preparation or at the tomography measurement. In
general we need to know the contributions $\chi^{\rm prep}$ and
$\chi^{\rm meas}$ from both errors separately because their addition
depends on the gate $U$ [Eq.\ (\ref{spam-exp})]. This can be done if
some high-fidelity single-qubit gates are available; then analyzing
the change of $\chi^{\rm err,exp}$ with application of different
high-fidelity gates, we can separate the contributions from
$\chi^{\rm prep}$ and $\chi^{\rm meas}$. Note that this method
assumes that the SPAM-errors can be represented as error processes
at the preparation and tomography stages; the accuracy of this
assumption is questionable. One of the ways to check this assumption
is to check one of its predictions: Eq.\ (\ref{spam-fid}) says that
the ``actual'' fidelity of a quantum gate is the ratio of its
QPT-measured fidelity and fidelity of the no-gate operation. The
gate fidelity  calculated in this way can then be compared with the
fidelity obtained from the randomized benchmarking.

    The appendices of this paper are to a significant extent
separated from the main text. In Appendix A we consider several
simple examples of $\chi$-matrices for unitary evolution and
decoherence, including the energy relaxation and pure dephasing
(Markovian and non-Markovian). In Appendix B we discuss unraveling
of the Lindblad-form evolution into the ``jump'' and ``no-jump''
scenarios, which can bring useful intuition in the analysis of
decoherence; several examples are considered to illustrate the
technique.

Reiterating the main point of this paper, we think that
characterization of quantum gates by error matrices in the Pauli
basis is a convenient way of presenting experimental QPT results.

\acknowledgements

The author thanks Yuri Bogdanov, Abraham Kofman, Justin Dressel,
Andrzej Veitia, Jay Gambetta, Michael Geller, and John Martinis for
useful discussions.
   The research was funded by the
Office of the Director of National Intelligence (ODNI), Intelligence
Advanced Research Projects Activity (IARPA), through the Army
Research Office Grant No. W911NF-10-1-0334. All statements of fact,
opinion, or conclusions contained herein are those of the authors
and should not be construed as representing the official views or
policies of IARPA, the ODNI, or the U.S. Government. We also
acknowledge support from the ARO MURI Grant No. W911NF-11-1-0268.

\appendix

\section{Simple examples of QPT}

In this Appendix we consider several simple examples of the quantum
processes, for which we calculate the standard process matrix $\chi$
of the QPT.

The matrix $\chi$ is defined via Eq.\ (\ref{chi-def}), which is
copied here for convenience,
    \be
    \rho_{\rm fin}= \sum\nolimits_{m,n} \chi_{mn} E_m
\rho_{\rm in}E_n^\dagger .
    \label{chi-def-2}\ee
For the operator basis  $\{ E_n\}$ we use the Pauli basis, so that
for one qubit it consists of four Pauli matrices,
 \begin{eqnarray}
&&  I=\left(\begin{array}{cc} 1&0\\0&1\end{array}\right) , \,\,\,
 X=\sigma_x=\left(\begin{array}{cc} 0&1\\1&0\end{array}\right) ,
  \label{Pauli-IX}
   \\
&& Y=\sigma_y=\left(\begin{array}{cc} 0&-i\\i&0\end{array}\right) ,
\,\,\,
 Z=\sigma_z=\left(\begin{array}{cc} 1&0\\0&-1\end{array}\right) ,
 \qquad
 \label{Pauli-YZ}\end{eqnarray}
while for several qubits the Kronecker (direct, outer, tensor)
product of these matrices is used. We use the notation \cite{N-C} in
which $\alpha
|0\rangle +\beta |1\rangle =\left(\begin{array}{c}\alpha \\
\beta\end{array} \right)$.

\subsection{Matrix $\chi$ for unitary operations}

\subsubsection*{One-qubit rotations}

As a very simple example, let us calculate the matrix $\chi$ for a
$Z$-rotation of a qubit over the angle $\varphi$. This realizes the
unitary operator $U$, which acts as $U (\alpha |0\rangle +\beta
|1\rangle)= \alpha |0\rangle +\beta e^{i\varphi} |1\rangle$ (we use
the sign convention of Ref.\ \cite{N-C}). Then since $|\psi\rangle
\langle \psi|\rightarrow U |\psi\rangle \langle \psi| U^\dagger $,
we have in general $\rho_{\rm fin}=U\rho_{\rm in}U^\dagger$. Let us
represent $U$ in the Pauli basis,
    \be
    U = \left(\begin{array}{cc}
    1&0\\0&e^{i\varphi}\end{array}\right) = e^{i\varphi/2}
    \left( (\cos \frac{\varphi}{2}) \, I -i(\sin \frac{\varphi}{2})\, Z\right),
    \label{U-Z}\ee
where the unimportant overall phase factor $e^{i\varphi/2}$ does not
affect the density matrix evolution, so that
    \be
\rho_{\rm fin} = [(\cos \frac{\varphi}{2}) I -i(\sin
\frac{\varphi}{2}) Z]
 \rho_{\rm in} [ (\cos \frac{\varphi}{2})  I +i(\sin \frac{\varphi}{2})
 Z]
    \label{rho-for-Z}\ee
(note that the Pauli matrices are Herimitian, so we do not need to
 conjugate them).
 Now by comparing Eq.\ (\ref{rho-for-Z}) with Eq.\ (\ref{chi-def-2})
 we immediately find
    \begin{eqnarray}
&&    \chi_{II} =\cos^2\frac{\varphi}{2}, \,\,\,
    \chi_{ZZ} =\sin^2\frac{\varphi}{2},
    \label{chi-Z-1}\\
&&    \chi_{IZ} =i\cos\frac{\varphi}{2} \sin\frac{\varphi}{2},
\,\,\,
    \chi_{ZI}=-\chi_{IZ},
    \label{chi-Z-2}\end{eqnarray}
other elements are zero. It is easy to see that the method of
finding $\chi$ by comparing Eq.\ (\ref{rho-for-Z}) with Eq.\
(\ref{chi-def-2}) is equivalent to using Eq.\ (\ref{chi-U}).

In a similar way we can calculate the matrix $\chi$ for a one-qubit
$X$-rotation over angle $\varphi$. The result is obviously the same
as Eqs.\ (\ref{chi-Z-1})--(\ref{chi-Z-2}), with index $Z$ replaced
by $X$:
  $\chi_{II} =\cos^2(\varphi /2)$, $\chi_{XX} =\sin^2 (\varphi /2)$,
$\chi_{IX}=-\chi_{XI}=i(\cos \varphi)/2$. Similarly, for
$Y$-rotations we replace $Z$ in Eqs.\
(\ref{chi-Z-1})--(\ref{chi-Z-2}) with $Y$.

    A qubit rotation about an axis $\vec{n}$ on the Bloch sphere
over an angle $\varphi$ corresponds to the unitary operator
\cite{N-C}
    \be
    U=\exp [ -i\frac{\varphi}{2} (\vec{n}\vec{\sigma})] =
    (\cos \frac{\varphi}{2})\, I - i  (\sin \frac{\varphi}{2}) \,
    (\vec{n}\vec{\sigma}),
    \ee
where $\vec{n}\vec{\sigma}\equiv n_x\sigma_x+ n_y\sigma_y+
n_z\sigma_z$. Note that this formula neglects the overall phase
factor (which does not exist in the Bloch-sphere space), as seen by
comparing it with Eq.\ (\ref{U-Z}).
 Using Eq.\ (\ref{chi-U}) or, alternatively, comparing equation
$\rho_{\rm fin}=U\rho_{\rm in}U^\dagger$ with the definition
(\ref{chi-def-2}), we still can easily find the elements of the
$\chi$-matrix; now all 16 elements are in general non-zero, though
they are determined by only 3 real parameters.

\subsubsection*{Two-qubit unitaries}

    Let us start with the case when the first qubit is $Z$-rotated
over the angle $\varphi$, while the second qubit is ``idling''. Then
the unitary operator is $U=[(\cos \frac{\varphi}{2}) \, I -i(\sin
\frac{\varphi}{2})\, Z] \otimes I$, and Eq.\ (\ref{rho-for-Z})
becomes
    \be
\rho_{\rm fin} = [(\cos I -i \sin Z)\otimes I]
 \rho_{\rm in} [ (\cos I +i \sin Z)\otimes I],
    \label{rho-for-Z-2}\ee
where for brevity we omit the argument $\varphi/2$ of sines and
cosines. Comparing this equation with (\ref{chi-def-2}), we have to
use double-letter combinations for both indices $m$ and $n$;
however, we see that the second-qubit letter is always $I$, so we
essentially obtain the single-qubit result
(\ref{chi-Z-1})--(\ref{chi-Z-2}) with added index $I$ for the second
qubit:
    \begin{eqnarray}
&&    \chi_{II,II} =\cos^2 (\varphi /2), \,\,\,
    \chi_{ZI,ZI} =\sin^2 (\varphi /2),
    \label{chi-Z-3}\\
&&    \chi_{II,ZI} = -\chi_{ZI,II}= i\cos (\varphi /2)\sin (\varphi
/2).
    \label{chi-Z-4}\end{eqnarray}

As another example, let us consider the  controlled-phase operation
(note that our use of the name ``controlled-phase'' is different
from the terminology of Ref.\ \cite{N-C}),
    \begin{eqnarray}
&& \hspace{-0.4cm}    U={\rm diag} (1,1,1,e^{i\theta })=
b(-ZZ+IZ+ZI)+c II, \, \qquad
    \\
&&  \hspace{-0.3cm}   b=(1-e^{i\theta})/4, \,\,\,
c=(3+e^{i\theta})/4,
    \end{eqnarray}
where in this (rather sloppy) notation we omit the Kronecker product
sign ``$\otimes$''. By comparing $U\rho_{\rm in} U^\dagger$ with
(\ref{chi-def-2}) we obtain 16 non-zero elements:
    \begin{eqnarray}
&&     \chi_{II,II}= |c|^2, \,\, \chi_{ZZ,ZZ}= \chi_{IZ,IZ}=
\chi_{ZI,ZI}= |b|^2, \, \qquad \,\,
    \label{c-p-1}\\
&& \chi_{IZ,ZI}= \chi_{ZI,IZ}= |b|^2,
  \\
&& \chi_{IZ,II}= \chi_{ZI,II}= bc^*, \,\,\,
 \chi_{II,IZ}= \chi_{II,ZI}= b^*c, \qquad
 \\
&& \chi_{ZZ,IZ}= \chi_{ZZ,ZI}=\chi_{IZ,ZZ}= \chi_{ZI,ZZ}= -|b|^2,
  \\
&& \chi_{ZZ,II}= -bc^*, \,\,\,
 \chi_{II,ZZ}= -b^*c .
    \label{c-p-5}\end{eqnarray}

     The controlled-phase gate becomes the CZ gate at
$\theta=\pi$. Then $b=c=1/2$, and all 16 elements of $\chi$ in Eqs.\
(\ref{c-p-1})--(\ref{c-p-5}) become $\pm 1/4$. Note that if in an
experimental realization the phase $\theta$ fluctuates symmetrically
around $\pi$ with a small variance $\langle
(\delta\theta)^2\rangle$, then the element $\chi_{II,II}$ increases
by $(3/16)\langle (\delta\theta)^2\rangle$, while other 15 elements
decrease in absolute value by $(1/16)\langle
(\delta\theta)^2\rangle$.

    For the perfect CNOT gate (with the first qubit being the control)
 the unitary can be represented as
    \be
    U=\frac{I+Z}{2}\otimes I +\frac{I-Z}{2}\otimes X,
    \label{U-CNOT}\ee
where we used the relations $(I+Z)/2=|0\rangle\langle 0|$ and
$(I-Z)/2=|1\rangle\langle 1|$ (here we again use the notation
``$\otimes$'' for more clarity). Then non-zero elements of $\chi$
are
    \begin{eqnarray}
&& \hspace{-0.4cm}     \chi_{II,II}=
\chi_{IX,IX}=\chi_{ZI,ZI}=\chi_{ZX,ZX}=1/4,
 \, \qquad
    \\
&& \hspace{-0.4cm}  \chi_{II,IX}=
\chi_{II,ZI}=\chi_{IX,II}=\chi_{ZI,II}=1/4 ,
  \\
&& \hspace{-0.4cm}  \chi_{II,ZX}= \chi_{ZX,II}=-1/4, \,\,
\chi_{IX,ZI}=\chi_{ZI,IX}=1/4, \,\, \qquad
 \\
&& \hspace{-0.4cm}  \chi_{IX,ZX}=
\chi_{ZI,ZX}=\chi_{ZX,IX}=\chi_{ZX,ZI}=-1/4 ,
    \end{eqnarray}
as directly follows from the combinations of 4 terms in Eq.\
(\ref{U-CNOT}).

    For the perfect $\sqrt{i{\rm SWAP}}$ gate the unitary is
    \begin{eqnarray}
&& \hspace{-0.3cm}    U=|00\rangle\langle 00| + |11\rangle\langle
11|
    -i (|01\rangle\langle 10| + |10\rangle\langle 01| ) \,\,\qquad
    \\
&&    = \frac{2+\sqrt{2}}{4} II
+\frac{2-\sqrt{2}}{4}ZZ-\frac{i\sqrt{2}}{4}(XX+YY), \qquad
    \end{eqnarray}
so the non-zero elements of the matrix $\chi$ are given by the
pairwise products of these 4 terms,
    \begin{eqnarray}
&& \hspace{-0.4cm}     \chi_{II,II}= (3+2\sqrt{2})/8, \,\,\,
\chi_{ZZ,ZZ}=(3-2\sqrt{2})/8,
 \, \qquad
    \\
&& \hspace{-0.4cm}  \chi_{XX,XX}= \chi_{YY,YY}=1/8 ,
  \\
&& \hspace{-0.4cm}
\chi_{XX,YY}=\chi_{YY,XX}=\chi_{II,ZZ}=\chi_{ZZ,II}=1/8,
 \\
&& \hspace{-0.4cm}
\chi_{II,XX}=\chi_{II,YY}=-\chi_{XX,II}=-\chi_{YY,II}
    \nonumber \\
&& \hspace{0.8cm} =i(\sqrt{2}+1)/8 ,
    \\
 && \hspace{-0.4cm}
\chi_{ZZ,XX}=\chi_{ZZ,YY}=-\chi_{XX,ZZ}=-\chi_{YY,ZZ}
    \nonumber \\
&& \hspace{0.8cm} =i(\sqrt{2}-1)/8 .
    \end{eqnarray}

\subsection{One-qubit decoherence}

\subsubsection*{Pure dephasing (exponential and non-exponential)}

  It is very easy to find the matrix $\chi$ for one qubit with pure
dephasing (assuming no other evolution). After waiting for time $t$,
the qubit is $Z$-rotated over a random angle $\varphi$. From the
definition $(\ref{chi-def-2})$ we see that for a random evolution we
simply need to average the $\chi$-matrix over the possible evolution
realizations. Therefore the $\chi$-matrix for pure dephasing is
given by averaging Eqs.\ (\ref{chi-Z-1})--(\ref{chi-Z-2}):
    \begin{eqnarray}
&& \hspace{-0.2cm}   \chi_{ZZ} =\langle \sin^2
\frac{\varphi}{2}\rangle =\frac{1-\langle \cos \varphi \rangle}{2},
\,\,\,\, \chi_{II} = 1 - \chi_{ZZ}, \qquad
    \label{deph-1}\\
&& \hspace{-0.2cm}    \chi_{IZ} =-\chi_{ZI}=i \langle \sin
\varphi\rangle /2,
    \label{deph-2}\end{eqnarray}
where $\langle ...\rangle$ denotes averaging over realizations. For
a symmetric probability density distribution of $\varphi$ we get
$\langle \sin \varphi \rangle =0$ and therefore
$\chi_{IZ}=\chi_{ZI}=0$, so the only non-zero elements are
$\chi_{II}$ and $\chi_{ZZ}$.

    It is important to emphasize that the result (\ref{deph-1})
does not assume exponential dephasing; it remains valid for an
``inhomogeneous'' contribution to the dephasing (slightly different
qubit frequencies in different experimental runs) and/or the ``1/f''
contribution (when the qubit frequency fluctuation has a broad range
of timescales). It is also important that the value $\langle \cos
\phi \rangle$ which determines $\chi_{ZZ}$ can be directly obtained
from the Ramsey-fringes data.

    Let us consider the Ramsey protocol: start with $|0\rangle$,
apply $\pi/2$ $X$-rotation, wait time $t$, apply the second $\pi/2$
rotation about the axis, which is shifted from $X$ by an angle
$\phi_R$ in the equatorial plane of the Bloch sphere, and finally
measure the probability $P(|1\rangle)$ of the state $|1\rangle$. It
is easy to find that for a pure dephasing (including non-exponential
case)
   \be
P(|1\rangle) =\frac{1}{2}+\frac{\langle \cos (\phi_R +\varphi
)\rangle }{2} =\frac{1}{2}+\frac{\langle \cos \varphi \rangle \cos
\phi_R }{2},
   \ee
where for the second equation we assumed $\langle \sin \varphi
\rangle =0$. In the case of exponential dephasing characterized by
the dephasing time $T_\varphi$, we have $\langle \cos \varphi
\rangle =\exp (-t/T_\varphi)$. In the general case the time
dependence of $\langle \cos \varphi \rangle$ is arbitrary; however,
it can be found experimentally from the amplitude of the Ramsey
oscillations and then can be used in Eq.\ (\ref{deph-1}) to obtain
$\chi_{ZZ}$.

    It is important to mention that experimentally $T_\varphi$ is
often defined as the time at which $\langle \cos \varphi \rangle
=e^{-1}$. If the qubit dephasing is due to fast (``white noise'')
fluctuations of the qubit energy, then $\langle \cos \varphi \rangle
=e^{-t/T_\varphi}$ and correspondingly
$\chi_{ZZ}=(1-e^{-t/T_\varphi})/2$, so that at short time, $t\ll
T_\varphi$, there is a linear dependence,
    \be
\chi_{ZZ}\approx t/2T_\varphi.
    \ee
 However, if the pure dephasing is
dominated by the very slow fluctuations of the qubit energy, then
$\langle \cos \varphi \rangle =\exp[-(t/T_\varphi)^2]$ and the
Ramsey-fringes dependence has a Gaussian shape. In this case at
$t\ll T_\varphi$ the dephasing error is quadratic in time,
    \be
\chi_{ZZ}\approx t^2/2T_\varphi^2 .
    \ee
In the presence of both mechanisms
    \be
    \langle \cos \varphi \rangle = \exp(-t/T_{\varphi, {\rm fast}})
     \exp[-(t/T_{\varphi, {\rm slow}})^2];
    \ee
this formula can also be used as an approximation in the case of a
broad range of the fluctuation timesclales. The corresponding
$\chi_{ZZ}$ is still given by Eq.\ (\ref{deph-1}).

 Note that in the presence of energy relaxation (discussed later)
the value of $\langle \cos \varphi \rangle$ can still be directly
extracted from the Ramsey-fringes data -- see Eq.\
(\ref{Ramsey-T1-deph}) below.

\subsubsection*{Energy relaxation}
\label{sec-ER}

    Now let us calculate the matrix $\chi$ taking into account the
qubit energy relaxation, but assuming the absence of pure dephasing.
Let us start with the zero-temperature case (relaxation to the state
$|0\rangle$ only). Using ``unraveling'' of the energy relaxation in
the same way as in Refs.\ \cite{T1-uncol} and \cite{Keane-2012} (see
also Appendix B), we may think about two probabilistic scenarios:
  \be
  \alpha |0\rangle +\beta |1\rangle \rightarrow \left\{
  \begin{array}{c} |0\rangle \,\,\, {\rm with \, prob.} \,\,
  P_r=|\beta|^2 (1-e^{-t/T_1})   \\
\displaystyle{ \frac{\alpha |0\rangle +\beta e^{-t/2T_1}
|1\rangle}{\sqrt{|\alpha|^2+ |\beta|^2 e^{-t/T_1}}} } \,\,\, {\rm
with \, prob.} \,\,
  1-P_r
   \end{array}
    \right.
  \ee
(in the case of no relaxation the state evolves due to the Bayesian
update). This corresponds to the technique of Kraus operators and
for the density matrix gives
    \be
    \rho_{\rm in} \rightarrow A_r  \rho_{\rm in} A_r^\dagger +
    A_{no}  \rho_{\rm in} A_{no}^\dagger ,
    \label{rel-Kraus-eq}\ee
where the Kraus operators $A_r$ (for the scenario with relaxation)
and $A_{no}$ (for the scenario with no relaxation) are
    \be
    A_r =\left(\begin{array}{cc}0 & \sqrt{1-e^{-t/T_1}} \\ 0 & 0
\end{array} \right) , \,\,\,
 A_{no} =\left( \begin{array}{cc} 1 & 0 \\ 0 & e^{-t/2T_1}
\end{array} \right) .
    \ee
Note that $A_r^\dagger A_r  +A_{no}^\dagger A_{no} =\openone$ (the
completeness relation). Here we use the standard Kraus-operator
representation \cite{N-C,Kraus-book}, in contrast to the somewhat
modified representation (\ref{Kraus-representation}).

    To find the $\chi$-matrix, we expand the Kraus operators in the
Pauli basis,
    \begin{eqnarray}
&& A_r= \sqrt{1-e^{-t/T_1}} \, \frac{X+iY}{2}, \,\,\,
    \\
&& A_{no}= \frac{1+e^{-t/2T_1}}{2} \, I + \frac{1-e^{-t/2T_1}}{2} \,
Z .
    \end{eqnarray}
Now comparing evolution (\ref{rel-Kraus-eq}) with the form
(\ref{chi-def-2}), we collect the $\chi$-matrix elements (the
relaxation term brings elements involving $X$ and $Y$, while the
no-relaxation term brings elements with $I$ and $Z$):
    \begin{eqnarray}
&& \hspace{-0.3cm} \chi_{XX}=\chi_{YY}=(1-e^{-t/T_1})/4 ,
  \\
&& \hspace{-0.3cm} \chi_{XY}=-\chi_{YX}=-i (1-e^{-t/T_1})/4 ,
   \\
&& \hspace{-0.3cm} \chi_{II}=(1+e^{-t/2T_1})^2 /4, \,\,
\chi_{ZZ}=(1-e^{-t/2T_1})^2 /4, \,\, \qquad
  \\
&& \hspace{-0.3cm} \chi_{IZ}=\chi_{ZI}= (1-e^{-t/T_1})/4 .
    \end{eqnarray}
Note that at small $t/T_1$ the element $\chi_{ZZ}$ is quadratic in
time (very small), while other elements (except  $\chi_{II}$) are
linear in time (in this case $\chi_{ZZ}\approx |\chi_{IZ}|^2$, as
expected from the discussion in Sec.\ \ref{sec-properties}). Also
note that the non-zero elements in the left column and top row
($\chi_{IZ}$ and $\chi_{ZI}$) are real and come from the
no-relaxation scenario (see Sec.\ \ref{sec-properties}).

    For a non-zero temperature there are two kinds of the relaxation
processes (up and down) with the rates $\Gamma_\uparrow$ and
$\Gamma_\downarrow$ satisfying the standard relations
$\Gamma_\uparrow +\Gamma_\downarrow =1/T_1$ and $\Gamma_\uparrow
/\Gamma_\downarrow =\exp (-E/T)$, where $T$ is temperature and
$E=E_1-E_0$ is the energy difference between the qubit states; this
gives $\Gamma_{\uparrow ,\downarrow}^{-1} = T_1 (1 + e^{\pm E/T})$.
Correspondingly, there are three scenarios with the Kraus operators
    \begin{eqnarray}
&& \hspace{-0.4cm}   A_{r\downarrow} =\left(\begin{array}{cc}0 &
\sqrt{1-e^{-\Gamma_\downarrow t}} \\ 0 & 0 \end{array} \right) ,
\,\, A_{r\uparrow} =\left(\begin{array}{cc}0 & 0 \\
\sqrt{1-e^{-\Gamma_\uparrow t}} & 0
\end{array} \right) \,\,  \qquad
   \\
&& \hspace{0.3cm} A_{no} =\left( \begin{array}{cc} e^{-\Gamma_\uparrow t/2} & 0 \\
0 & e^{-\Gamma_\downarrow t/2}
\end{array} \right) .
    \end{eqnarray}
Then in a similar way as above we find the $\chi$-matrix elements:
    \begin{eqnarray}
&& \chi_{XX}=\chi_{YY}=(1-e^{-\Gamma_\downarrow t})/4 +
(1-e^{-\Gamma_\uparrow t})/4  , \qquad
    \label{rel-temp-1}  \\
&& \chi_{XY}=-\chi_{YX}=-i (e^{-\Gamma_\uparrow t}
-e^{-\Gamma_\downarrow t})/4 ,
   \\
&& \chi_{II}=(e^{-\Gamma_\uparrow t/2 }+e^{-\Gamma_\downarrow
t/2})^2 /4, \,\,
   \\
&&  \chi_{ZZ}= (e^{-\Gamma_\uparrow t/2 }-e^{-\Gamma_\downarrow
t/2})^2 /4, \qquad
  \\
&& \chi_{IZ}=\chi_{ZI}= (e^{-\Gamma_\uparrow t}
-e^{-\Gamma_\downarrow t})/4 .
    \label{rel-temp-5}    \end{eqnarray}

\subsubsection*{Pure dephasing combined with energy relaxation}

    Phase evolution commutes with the energy relaxation, therefore we
may apply $Z$-rotation over a random angle $\varphi$ after the
energy relaxation. The $Z$-rotation does not affect scenario(s) with
relaxation, so we need to change only the Kraus operator for the
no-relaxation scenario: $A_{no}\rightarrow (I\cos \frac{\varphi}{2}
-iZ \sin \frac{\varphi}{2}) A_{no} $, then calculate the
corresponding elements of the $\chi$-matrix in the same way as above
and average over $\varphi$. Therefore, the elements $\chi_{XX}$,
$\chi_{YY}$, $\chi_{XY}$, $\chi_{YX}$ due to energy relaxation will
not be affected by the pure dephasing (since they come from the
relaxation scenario). The calculation shows that the elements
$\chi_{IZ}$ and $\chi_{ZI}$ are also not affected when $\langle \sin
\varphi \cos\varphi \rangle =0$ (satisfied for a symmetric noise),
so the only affected elements are $\chi_{ZZ}$ and $\chi_{II}$, which
are non-zero for both decoherence mechanisms. Using again the
condition $\langle \sin \varphi \cos\varphi \rangle =0$, it is easy
to find
    \begin{eqnarray}
&& \chi_{ZZ}= \chi_{ZZ}^{deph}\chi_{II}^{rel} +
\chi_{ZZ}^{rel}\chi_{II}^{deph} ,
    \label{deph+rel-ZZ}   \\
&& \chi_{II}= \chi_{II}^{deph}\chi_{II}^{rel} +
\chi_{ZZ}^{deph}\chi_{ZZ}^{rel} ,
    \end{eqnarray}
where $\chi^{deph}$ and $\chi^{rel}$ correspond to pure dephasing
and energy relaxation, and were calculated above [Eqs.\
(\ref{deph-1}) and (\ref{rel-temp-1})--(\ref{rel-temp-5})]. For
completeness let us also show the unaffected elements:
    \begin{eqnarray}
&& \chi_{XX}= \chi_{YY}= \chi_{XX}^{rel},
  \,\,\, \chi_{XY}= -\chi_{YX} =\chi_{XY}^{rel} , \qquad
   \\
&& \chi_{IZ}= \chi_{ZI} =\chi_{IZ}^{rel} .
    \label{deph+rel-IZ}\end{eqnarray}

Another way of deriving Eqs.\
(\ref{deph+rel-ZZ})--(\ref{deph+rel-IZ}) is the following. Let us
apply pure dephasing after energy relaxation and write the
composition of quantum operations as
   \be
    \rho_{\rm fin}= \sum_{m,n} \sum_{m'n'} \chi_{mn}^{rel} \chi_{m'n'}^{deph}
E_{m'}  E_m \rho_{\rm in}E_n^\dagger E_{n'}^\dagger .
    \label{chi-composition}\ee
Even though a product of Pauli matrices is a Pauli matrix (possibly
with a phase factor) and therefore $\chi$-matrix of a composition of
operations can in principle be calculated in a straightforward way,
usually this is a  very cumbersome procedure. However, in our case
the matrix $\chi^{deph}$ has only two non-zero elements
($\chi^{deph}_{II}$ and $\chi^{deph}_{ZZ}$), so the calculation is
not very long and leads to Eqs.\
(\ref{deph+rel-ZZ})--(\ref{deph+rel-IZ}).

    Now when we have explicit formulas for the $\chi$-matrix elements,
which depend on $t/T_1$, temperature, and $\langle \cos \varphi
\rangle$, let us discuss again how to extract $\langle \cos \varphi
\rangle$ from the Ramsey-fringes data. In the presence of energy
relaxation (at arbitrary temperature) the Ramsey oscillations are
   \begin{eqnarray}
&& P(|1\rangle) =\frac{1}{2}+\frac{e^{-(\Gamma_\downarrow
+\Gamma_\uparrow) t/2}}{2} \, \langle \cos (\phi_R+\varphi )\rangle
 \qquad
    \\
&& \hspace{1.1cm} = \frac{1}{2}+ \frac{1}{2} \, e^{-t/2T_1} \langle
\cos \varphi \rangle \, \cos \phi_R .
   \label{Ramsey-T1-deph}\end{eqnarray}
   Therefore, if the energy relaxation time $T_1$ is
measured separately, the Ramsey data give the value of $\langle \cos
\varphi \rangle$ at any time $t$. This value can be used to
calculate the $\chi$-matrix even in the case of arbitrary
non-exponential dephasing.

    Now let us discuss the $\chi$-matrix at a relatively short time
$t$ and neglect the terms quadratic in time. Then  we obtain
    \begin{eqnarray}
&& \chi_{XX}=\chi_{YY}= t/4T_1 , \,\,\, \chi_{ZZ}=(1-\langle \cos
\varphi \rangle )/2, \qquad
  \label{chi-r+d-1}\\
  &&  \,\,
 \chi_{II}= 1- t/2T_1 - (1-\langle \cos \varphi \rangle )/2,  \qquad
 \label{chi-r+d-2}\\
&& \chi_{XY}=-\chi_{YX}=-i (t/4T_1) \tanh(E/2T) ,
  \label{chi-r+d-3}\\
&& \chi_{IZ}=\chi_{ZI}= (t/4T_1) \tanh(E/2T).
    \label{chi-r+d-4}\end{eqnarray}
In the case of exponential pure dephasing $\langle \cos\varphi
\rangle = \exp (-t/T_\varphi)$,  and then $\chi_{ZZ}=t/2T_\varphi$,
where $T_\varphi^{-1}=T_2^{-1}- (2T_1)^{-1}$ and $T_2$ is the
dephasing time.

    \section{Interpretation of the Lindblad-form master equation}

In this Appendix we discuss the technique of Kraus operators applied
to the evolution described by the standard Lindblad-form master
equation. We show that each Lindblad term describes two evolutions:
a ``jump'' process with some rate and a continuous evolution between
the jumps caused by the absence of jumps. Such interpretation can be
useful in intuitive analysis of decoherence processes.

    A Markovian  evolution of a quantum system is usually described
by the Lindblad-form master equation [for convenience we copy Eq.\
(\ref{Lindblad-evol}) here]
    \be
    \dot\rho = - \frac{i}{\hbar} [H,\rho] + \sum_n \Gamma_n ( B_n\rho B_n^\dagger
-\frac{1}{2} B_n^\dagger B_n\rho -\frac{1}{2} \rho B_n^\dagger B_n
),
    \label{Lindblad}\ee
where the first term describes the unitary evolution due to the
Hamiltonian $H$, while the $n$th decoherence mechanism is described
by the ``rate'' $\Gamma_n$ (real number with dimension s$^{-1}$) and
dimensionless operator $B_n$. Note that mathematically $\Gamma_n$
can be absorbed by redefining $\tilde{B}_n=\sqrt{\Gamma_n}\, B_n$;
however, we do not do this because $\Gamma_n$ and $B_n$ have
separate physical meanings.

Our goal here is to discuss a simple physical interpretation of the
decoherence terms in the Lindblad equation. For simplicity let us
neglect the unitary evolution ($H=0$) and consider first only one
decoherence mechanism; then we can omit the index $n$ (summation
over $n$ is simple).

 It is easy to check that the term $\Gamma
B\rho B^\dagger$ corresponds to the abrupt change (``jump'') of the
state
    \be
    |\psi \rangle \rightarrow \frac{B  |\psi \rangle}{\rm Norm}
    =\frac{B  |\psi \rangle}{|| B  |\psi \rangle ||}
    , \,\,\,\,\,\, \rho \rightarrow \frac{B \rho B^\dagger}{\rm
    Norm} = \frac{B \rho B^\dagger}{{\rm
    Tr} (B \rho B^\dagger)},
    \label{jump}\ee
which occurs with the rate (jump probability per second)
    \be
    \frac{dP}{dt}= \Gamma \, || B|\psi\rangle ||^2, \,\,\,\,
    \frac{dP}{dt}=
\Gamma \, {\rm Tr} ( B \rho B^\dagger) = \Gamma \, {\rm Tr} (
B^\dagger B \rho ).
    \label{rate}\ee
(Here we show the formulas in both the wavefunction and density
matrix languages; the wavefunction language is usually more
convenient to use.) Note a possible confusion in terminology: both
$\Gamma$ and $dP/dt$ are rates; to distinguish them let us call
$\Gamma$ the ``process rate'' or just ``rate'', while $dP/dt$ will
be called ``jump rate''.

    The remaining term $-(\Gamma/2)(B^\dagger B\rho + \rho B^\dagger
B)$ in the Lindblad form corresponds to the jump process
(\ref{jump}) {\it not happening}. The physical reason of this
evolution ``when nothing happens'' is the same as the partial
collapse in the null-result measurement \cite{Molmer-92,Katz-06}:
this is essentially the Bayesian update \cite{Kor-Bayes}, which
accounts for the information that the jump did not happen.
Therefore, {\it the physical meaning of the Lindblad form is the
description of two scenarios: the jump process either happening or
not happening.}

    To see this mathematically, let us use the technique of Kraus
operators \cite{N-C,Kraus-book}, in which the evolution $\rho_{\rm
in}\rightarrow \rho_{\rm fin}$ is unraveled into the probabilistic
mixture of ``scenarios'' described by Kraus operators $A_k$,
    \be
    |\psi_{\rm in} \rangle \rightarrow \frac{A_k
|\psi_{\rm in}\rangle }{\rm
    Norm}, \,\,\, \rho_{\rm in}\rightarrow \frac{A_k \rho_{\rm in}
    A_k^\dagger}{\rm Norm} ,
   \ee
with probabilities $P_k={\rm Tr} (A_k^\dagger A_k \rho_{\rm in})$
[in the wavefunction language $P_k=||A_k |\psi_{\rm in}\rangle
||^2={\rm Tr} (A_k^\dagger A_k |\psi_{\rm in}\rangle \langle
\psi_{\rm in}|)$]. The sum of the probabilities should be equal to
1, which leads to the completeness relation
    \be
    \sum\nolimits_{k} A_k^\dagger A_k =\openone.
     \label{completeness}\ee
Note the change of notations compared with Eq.\
(\ref{Kraus-representation}).

 During a short time $\Delta t$ the evolution (\ref{Lindblad})
can be described by two scenarios: the jump either occurs or not,
with the corresponding Kraus operators $A_{jump}$ and $A_{no}$. For
the jump scenario
    \be
A_{jump}=\sqrt{\Gamma \Delta t}\, B ,
    \ee
so that $A_{jump} \rho A_{jump}^\dagger = (\Gamma \Delta t) B\rho
B^\dagger$ gives the same contribution to $\Delta \rho=\rho
(t+\Delta t)-\rho (t)$ as the term $\Gamma B\rho B^\dagger$ in Eq.\
(\ref{Lindblad}). The no-jump Kraus operator $A_{no}$ should satisfy
the completeness relation (\ref{completeness}), $A_{jump}^\dagger
A_{jump} +A_{no}^\dagger A_{no}=\openone$. Using the Bayesian-update
approach \cite{Kor-Bayes} in which $A_{no}^\dagger =A_{no}$, we find
    \begin{eqnarray}
&&    A_{no}=\sqrt{\openone -A_{jump}^\dagger A_{jump}}\approx
    \openone -\frac{1}{2} \, A_{jump}^\dagger A_{jump} \qquad
        \label{A-no}\\
&& \hspace{0.7cm} = \openone -\frac{1}{2}\, \Gamma \Delta t \,
B^\dagger B.
    \label{A-no-2}\end{eqnarray}
Note that $A_{jump}^\dagger A_{jump}$ is a positive Hermitian
operator and $\openone - A_{jump}^\dagger A_{jump}$ is also a
positive operator. A square root of a positive operator is defined
via taking square roots of its eigenvalues in the diagonalizing
basis. This is why in Eq.\ (\ref{A-no}) we deal with operators
essentially as with numbers.

Using Eq.\ (\ref{A-no-2}) for $A_{no}$, we find in the linear
    order
    \be
    A_{no}\, \rho A_{no}^\dagger \approx \rho - \frac{1}{2}\,
    \Gamma \Delta
    t\, B^\dagger B \, \rho -\frac{1}{2}\, \Gamma \Delta t\, \rho
     B^\dagger B ,
    \label{A-no-evol}\ee
with this linear-order approximation becoming exact at $\Delta
t\rightarrow 0$. It is easy to see that Eq.\ (\ref{A-no-evol}) gives
the evolution described by the term $-(\Gamma/2)(B^\dagger B\rho +
\rho B^\dagger B)$ in the Lindblad form (\ref{Lindblad}).

Thus we have shown that the Lindblad-form master equation
(\ref{Lindblad}) with one decoherence term describes a jump process
$B$ [see Eq.\ (\ref{jump})] occurring with the jump rate
(probability per second) $\Gamma \,{\rm Tr}(B\rho B^\dagger)$. In
the case of several decoherence mechanisms there are several Kraus
operators $A_{n, jump}=\sqrt{\Gamma_n\Delta t}\, B_n$ describing the
jumps $|\psi\rangle \rightarrow B_n |\psi\rangle /{\rm Norm}$ during
a short duration $\Delta t$. The no-jump Kraus operator in this case
is $A_{no}=\sqrt{\openone -\sum_n (A_{n,jump}^\dagger A_{n,jump})}
\approx \openone -\sum_n (\frac{1}{2}\Gamma_n \Delta t\, B_n^\dagger
B_n)$, which contributes to all terms $-(\Gamma_n/2)(B_n^\dagger
B_n\rho + \rho B_n^\dagger B_n)$ in Eq.\ (\ref{Lindblad}).

    Note that  Eq.\ (\ref{A-no}) and its generalization for
several processes is not the unique form for $A_{no}$, which follows
from the completeness relation (\ref{completeness}). It is formally
possible to add an arbitrary unitary rotation $U$, so that
$A_{no}=U\sqrt{\openone -\sum_n (A_{n,jump}^\dagger A_{n,jump})}$,
which does not affect $A_{no}^\dagger A_{no}$. Using a natural
assumption that $U\rightarrow \openone$ at $\Delta t\rightarrow 0$,
we can expand $U$ in the linear order as $U=\openone -iH_a\Delta t$,
where $H_a$ should be a Hermitian matrix. Then $A_{no}\rho
A_{no}^\dagger$ acquires the extra term $-i[H_a,\rho]\Delta t$, from
which we see that $H_a$ is essentially an addition to the
Hamiltonian $H$. So the formalism permits a change of the
Hamiltonian due to decoherence, and therefore $H$ in Eq.\
(\ref{Lindblad}) should be considered as the effective Hamiltonian
(which may include the ``Lamb shift'' mechanism).

    Now let us consider several examples of decoherence
processes in the language of ``jumps''.

    \subsection{Energy relaxation of a qubit}

    For a qubit relaxation from the excited state $|1\rangle$ to the
ground state $|0\rangle$ with the rate $\Gamma = 1/T_1$, the jump
operator is $B=\left(\begin{array}{cc}0 & 1 \\ 0 & 0
\end{array} \right)$, where the ground state corresponds to the
upper line, so that $B|1\rangle =|0\rangle$, $B|0\rangle =0$. In
this case $B^\dagger B=\left(\begin{array}{cc}0 & 0 \\ 0 & 1
\end{array} \right)$, so that the no-jump evolution during time
$\Delta t$ changes the qubit state as
    \be
\alpha |0\rangle +\beta |1\rangle \rightarrow \frac{\alpha |0\rangle
+\beta e^{-\Delta t/2T_1} |1\rangle }{\rm Norm} \approx \frac{\alpha
|0\rangle +\beta (1-\frac{\Delta t}{2T_1}) |1\rangle }{\rm Norm},
    \label{qubit-relax-no-jump}\ee
while in the case of jump obviously $\alpha |0\rangle +\beta
|1\rangle \rightarrow |0\rangle$ (the jump rate is $\Gamma
|\beta|^2$). These two evolutions give correct density matrix, which
also follows from the Lindblad equation solution (see Sec.\ IIA of
Ref.\ \cite{Keane-2012} for more detailed discussion).

    The excitation processes  $|0\rangle \rightarrow |1\rangle$ can be
taken into account in the similar way using an additional
Lindblad-form term.

    \subsection{Pure dephasing of a qubit}

   The physical mechanism of the Markovian pure dephasing in
superconducting qubits is the fast (``white noise'') fluctuations of
the qubit energy, which leads to random fluctuations of the qubit
phase $\varphi = \arg (\alpha^*\beta)=\arg (\rho_{01})$, so that the
random phase shift $\Delta\varphi$ accumulated during a short time
$\Delta t$ has the Gaussian probability distribution with zero mean
and variance $\overline{(\Delta\varphi)^2}=2\Delta t/T_\varphi$,
where $T_\varphi$ is the dephasing time.

It is easy to check that the correct evolution due to pure dephasing
[$\rho_{01}(t)=\rho_{01}(0)\, e^{-t/T_\varphi }$,
$\rho_{00}(t)=\rho_{00}(0)$, $\rho_{11}(t)=\rho_{11}(0)$] can be
reproduced using the Lindblad equation (\ref{Lindblad}) with $\Gamma
= 1/2T_\varphi$ and $B=\left(\begin{array}{cc}1 & 0 \\ 0 & -1
\end{array} \right)$. This means that instead of the physically
correct picture of continuous change of $\varphi$, we may use a
completely different picture: random jumps of the phase $\varphi$ by
$\pi$,
    \be
\alpha |0\rangle +\beta |1\rangle \rightarrow \alpha |0\rangle
-\beta |1\rangle ,
    \ee
occuring with the jump rate [see Eq.\ (\ref{rate})] $\Gamma \, {\rm
Tr} (B^\dagger B\rho)=1/2T_\varphi$, which in this case is
independent of the qubit state and equal to $\Gamma$. Note that
$B^\dagger B=\openone$, so in this case there is no no-jump
evolution.

    Since both pictures lead to the same evolution of $\rho (t)$, we
can use any of them, depending on convenience in a particular
problem.
   One more picture which can be used for pure dephasing is
the ``random measurement of state $|0\rangle$'', for which
$B=\left(\begin{array}{cc}1 & 0 \\ 0 & 0
\end{array} \right)$ and $\Gamma =2/T_\varphi$. It is easy to see
that it leads to the same Lindblad equation, but has different jump
process and non-trivial no-jump evolution. Obviously, we can also
use $B=\left(\begin{array}{cc}0 & 0 \\ 0 & 1
\end{array} \right)$ and $\Gamma =2/T_\varphi$ for the same
master equation (see the brief discussion in Sec.\
\ref{sec-Lindblad-error} of the transformation $B\rightarrow
B-b\openone$ in the Lindblad equation).

    \subsection{Resonator state decay}

    The decay of a resonator state is usually characterized by the energy
decay rate $\kappa$, so that in the qubit terminology
$\kappa=1/T_1$. The standard Lindblad form (\ref{Lindblad}) in this
case has $\Gamma=\kappa$ and $B=a$, where $a$ is the annihilation
operator, $a|n\rangle =\sqrt{n}\,|n-1\rangle$. The jump evolution is
then
    \be
    |\psi \rangle \rightarrow \frac{a |\psi \rangle}{\rm Norm},
    \label{res-jump}\ee
with the jump rate [see Eq.\ (\ref{rate})] $\kappa \, {\rm
Tr}(a^\dagger a \rho)=\kappa \, \overline{n}$, where $\overline{n}$
is the average number of photons. The no-jump evolution during time
$\Delta t$ is then
    \be
    \sum_n \alpha_n \, |n\rangle \rightarrow
    \frac{\sum_n e^{-n\kappa \Delta t/2} \alpha_n  |n\rangle }{\rm Norm}
    \approx
     \frac{\sum_n (1-n\frac{\kappa}{2} \Delta t ) \alpha_n |n\rangle }
     {\rm Norm},
    \label{res-no-jump}\ee
which is similar to the no-jump evolution
(\ref{qubit-relax-no-jump}) for the qubit. Note that the probability
of the no-jump evolution is given by the squared norm, ${\rm
Norm}^2\approx 1- \kappa \, \overline{n} \Delta t$, so that the sum
of the jump and no-jump probabilities is 1.

    It is interesting to analyze the special case: evolution of
a coherent state, $|\lambda\rangle \equiv e^{-|\lambda |^2/2}\sum_n
(\lambda^{n}/\sqrt{n!})|n\rangle$. Since it is the eigenstate of the
operator $a$, the jump evolution (\ref{res-jump}) does not change
the state, while the continuous no-jump evolution essentially
changes the parameter $\lambda$ in time, $\lambda (\Delta
t)=e^{-\kappa\Delta t/2} \lambda (0)$. Therefore the energy decay is
due to the no-jump evolution only. In this (very unusual) case a
pure initial state remains pure because the jump does not change the
state and therefore it is essentially a one-scenario (no-jump)
evolution.

    The standard Lindblad-form evolution for the resonator with
$\Gamma=\kappa$ and $B=a$ can be understood in the following way.
Let us assume that $\kappa$ is the coupling with the outside modes,
and let us imagine (gedankenexperiment) that we use an ideal photon
detector, which clicks every time when a photon escapes from the
resonator. When the detector clicks, we know that there is one
photon less in the resonator. However, all transitions $|n\rangle
\rightarrow |n-1\rangle$ are indistinguishable (because the energy
levels are equidistant), and therefore the Bayesian update of the
quantum state \cite{Kor-Bayes} involves only the rates, which are
proportional to $n$,
    \be
\alpha_n |n\rangle \rightarrow \frac{\alpha_n \sqrt{\kappa\, n} \,
|n-1\rangle }{\rm Norm}.
    \ee
It is easy to see that this Bayesian update coincides with the jump
process (\ref{res-jump}), and therefore the Lindblad-form evolution
can be interpreted as being due to an outside measurement with a
single-photon detector.
  [Note that the Lindblad-form master equation describes the
ensemble-averaged evolution; this is why we are free to choose any
measurement model, in contrast to the case of individual (selective)
evolution, which depends on what is actually measured.]

    \subsection{Energy relaxation in a 3-level qubit}

    The energy relaxation (at zero temperature) in a slightly
ahnarmonic superconducting 3-level qubit is sometimes described by
essentially the same Lindblad equation as for the resonator, with
$\Gamma=1/T_1$ and $B=a$, so that $B|2\rangle =\sqrt{2} \,
|1\rangle$, $B|1\rangle =|0\rangle$,  $B|0\rangle =0$. This is
actually incorrect. The reason is that the qubit is anharmonic, and
therefore in the described above gedankenexperiment it is in
principle possible to distinguish transitions $|2\rangle\rightarrow
|1\rangle$ and $|1\rangle\rightarrow |0\rangle$. Therefore the
energy relaxation should be described by two Lindblad terms, with
    \begin{eqnarray}
    \Gamma_{1\rightarrow 0}=\frac{1}{T_1}, \,\,\,
    B_{1\rightarrow 0}= \left(\begin{array}
    {ccc}0 & 1 & 0\\ 0 & 0 &0 \\ 0 & 0 & 0
\end{array} \right),
    \label{3-level-1}\\
     \Gamma_{2\rightarrow 1}\approx \frac{2}{T_1}, \,\,\,
  B_{2\rightarrow 1}= \left(\begin{array}
    {ccc}0 & 0 & 0\\ 0 & 0 &1 \\ 0 & 0 & 0
\end{array} \right) ,
    \label{3-level-2}\end{eqnarray}
where the top row corresponds to the state $|0\rangle$ and the
bottom row corresponds to $|2\rangle$. The formula
$\Gamma_{2\rightarrow 1}\approx 2/T_1$ is not exact because of
anharmonicity.
 If for more accuracy we want
to take into account direct transitions $|2\rangle \rightarrow
|0\rangle$ (which may become allowed because of anharmonicity), we
can introduce the third term in the similar way.

    Compared with the two-process description
(\ref{3-level-1})--(\ref{3-level-2}), the (incorrect) one-process
description (with $B=a$) adds the extra term $(\sqrt{2}\,/T_1)
\rho_{12}$ into $\dot{\rho}_{01}$, while other terms in these two
approaches coincide.

\vspace{0.2cm}

    Concluding this Appendix, we emphasize that unraveling of
the Lindblad-form evolution into the ``jump'' and ``no-jump''
processes is often useful for gaining intuition in the analysis of
decoherence, and it may also be useful in analytical and numerical
calculations.


\begin{thebibliography}{99}
\bibitem{Frobenius} To discuss orthogonality and normalization of
matrices, we need to define the inner product of matrices $A$ and
$B$ of equal size. This can be done by direct generalization of the
innner product of two vectors: the matrices are treated as vectors,
and the inner product is defined element-by-element as $\langle
A|B\rangle \equiv \sum_{i,j} A^*_{ij}B_{ij}$. This is called
Frobenius or Hilbert-Schmidt inner product; we use the notation
similar to the notation for vectors in quantum mechanics, while in
mathematics notations $B\! :\! A$ or $\langle B,A\rangle$ are often
used. It is easy to see that $\langle
A|B\rangle=\sum_{i,j}(A^\dagger)_{ji}B_{ij}={\rm Tr}(A^\dagger B)$.
This definition leads to the Frobenius (Hilbert-Schmidt) norm for a
matrix: $||A||=\sqrt{{\rm Tr}(A^\dagger
A)}=\sqrt{\sum_{i,j}|A_{ij}|^2}$.
\end{thebibliography}
\end{document}